\documentclass[prl,showpacs,twocolumn]{revtex4-1}
\usepackage{amsmath,amsfonts,amssymb,bm}				
\allowdisplaybreaks[1]
\usepackage[mathscr]{eucal}
\usepackage{nicefrac}														
\usepackage[T1]{fontenc}												
\usepackage{dsfont}														
\usepackage{graphicx}
\usepackage{subfigure}
\usepackage{float}															
\usepackage{wrapfig}
\usepackage{sidecap}														
\usepackage{placeins}
\usepackage{hyperref}
\newcounter{myequation}
\makeatletter
\@addtoreset{equation}{myequation}
\makeatother
\newcounter{myequation2}
\makeatletter
\@addtoreset{figure}{myequation2}
\makeatother

\usepackage{color}

\newcommand{\ma}[1]{\begin{align}{#1}\end{align}}

\newcommand{\bra}[1]{\langle #1 |}
\newcommand{\ket}[1]{|#1\rangle}

\begin{document}
\title{Observing the Nonequilibrium Dynamics of the Quantum Transverse-Field Ising Chain in Circuit QED}
\author{Oliver Viehmann,$^1$ Jan von Delft,$^1$ and Florian Marquardt$^2$}
\affiliation{$^1$Physics Department,
             Arnold Sommerfeld Center for Theoretical Physics,
             and Center for NanoScience,\\
             Ludwig-Maximilians-Universit\"at,
             Theresienstra{\ss}e 37,
             80333 M\"unchen, Germany}
\affiliation{$^2$Institut for Theoretical Physics, Universit\"at Erlangen-N\"urnberg, Staudtstra\ss e 7, 91058 Erlangen, Germany}

\begin{abstract}
We show how a quantum Ising spin chain in a time-dependent transverse magnetic field can be simulated and experimentally probed in the framework of circuit QED with current technology. The proposed setup provides a new platform for observing the nonequilibrium dynamics of interacting many-body systems. We calculate its spectrum to offer a guideline for its initial experimental characterization. We demonstrate that quench dynamics and the propagation of localized excitations can be observed with the proposed setup and discuss further possible applications and modifications of this circuit QED quantum simulator.

\end{abstract}

\maketitle
The promising idea of tackling complex quantum many-body problems by quantum simulations \cite{Feynman1982,Buluta2009} has become even more compelling recently, due to the widespread current interest in nonequilibrium dynamics. Indeed, experiments with cold atoms in optical lattices \cite{Jaksch1998,Greiner2002,Simon2011,Trotzky2012} and ions \cite{Friedenauer2008,Islam2011,Lanyon2011,Britton2012} have already made impressive progress in this regard. At the same time, the capabilities of scalable, flexible solid-state platforms are developing rapidly. In particular, circuit quantum electrodynamics (cQED) architectures of superconducting artificial atoms and microwave resonators \cite{Wallraff2004,McDermott2005,DiCarlo2010,Vijay2011,Johnson2011,Mariantoni2011,Fedorov2012,Dewes2012,Reed2012} are now moving toward multiatom, multiresonator setups with drastically enhanced coherence times, making them increasingly attractive candidates for quantum simulations \cite{Houck2012}. Here we propose and analyze a cQED design that simulates a quantum transverse-field Ising chain with current technology. Our setup can be used to study quench dynamics, the propagation of localized excitations, and other nonequilibrium features, in a field theory exhibiting a quantum phase transition (QPT) \cite{Sachdev1999}, and based on a design that could easily be extended to break the integrability of the system.

The present Letter takes a different path than the proposals for simulating Bose-Hubbard-type many-body physics in cavity arrays, which might be also realizable in cQED \cite{Greentree2006,Hartmann2006,Koch2009,Schiro2012,Hwang2012,Houck2012}. It is based on a possibly simpler concept -- direct coupling of artificial atoms -- that naturally offers access to quantum magnetism. The transverse-field Ising chain (TFIC) is a paradigmatic quantum many-body system. It is exactly solvable \cite{Lieb1961,Pfeuty1970} and thus serves as a standard theoretical example in the context of nonequilibrium thermodynamics and quantum criticality \cite{Sachdev1999,Polkovnikov2011,Barouch1970,Igloi2000,Calabrese2006_2007,Igloi2011,Sachdev1997}. Our proposal to simulate the TFIC and its nonequilibrium dynamics might help to mitigate the lack of experimental systems for testing these results. Moreover, the experimental confirmation of our predictions for various nonequilibrium scenarios in this integrable many-body system would serve as an important benchmark and allow one to proceed to variations of the design that break integrability or introduce other features.

\begin{figure}[b!]
	\centering
	\includegraphics[width=\columnwidth]{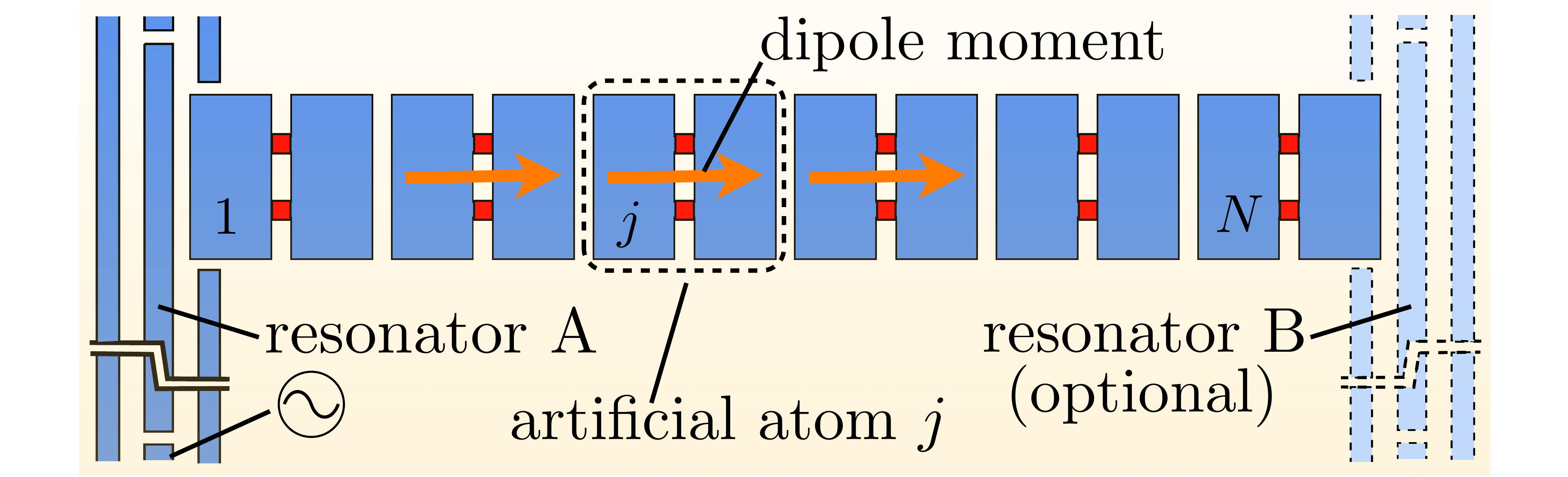}
	\caption{Circuit QED implementation of the Ising model with a transverse magnetic field. 
	The dipole moments of the artificial atoms tend to align.
	Resonator A (B) facilitates initialization and readout of the first ($N$th) artificial atom by standard circuit QED techniques.}
	\label{fig:1}
\end{figure}

\paragraph*{Implementation of the TFIC.---}A charge-based artificial atom (such as the Cooper-pair box or the transmon) \cite{Clarke2008} in a superconducting microwave resonator can be understood as an electric dipole (with dipole operator $\sigma_x$) that couples to the quantized electromagnetic field in the resonator \cite{Schoelkopf2008}. Consider the system of Fig.\ \ref{fig:1}, at first, without resonator B. 
\noindent
Only the first artificial atom couples to resonator A. However, all atoms couple directly (not mediated by a quantized field) to their neighbors via dipole-dipole coupling $\propto \sigma_x^i \sigma_x^j$ (for details, see Ref.\ \cite{Supplement}). Coupling of this type has already been demonstrated with two Cooper-pair boxes \cite{Pashkin2003} and two transmons \cite{Dewes2012}. Since this interaction is short ranged, we model our system by
\ma{
\mathcal{H} = \omega_0(a^\dagger a + 1/2) + g(a^\dagger + a)\sigma_x^1 + \mathcal{H}_I, \label{eq:full_model}
}
where $\mathcal{H}_I$ is the Hamiltonian of the TFIC,
\ma{
\mathcal{H}_I = \dfrac{\Omega}{2} \sum_{j=1}^N \sigma_z^j -J \sum_{j=1}^{N-1} \sigma_x^j \sigma_x^{j+1}. \label{eq:TFIC}
}
Here, $a^\dagger$ generates a photon with frequency $\omega_0$, and $\sigma_{x/z}^j$ is a Pauli matrix. That is, we consider the artificial atoms as two-level systems (qubits). This is justified even for weakly anharmonic transmons since the experiments proposed below involve only low atomic excitation probabilities or well controllable excitation techniques ($\pi$ pulses). Qubit 1 and the resonator couple with strength $g$. The qubit level spacing $\Omega>0$ is tunable rapidly ($\sim 1$ns) via the magnetic flux through the qubits' SQUID loops \cite{Wallraff2004,DiCarlo2010,Fedorov2012,Reed2012}. It corresponds to the transverse magnetic field in the usual TFIC. In our geometry, the qubit-qubit coupling strength $J$ is positive (ferromagnetic; the antiferromagnetic coupling $J<0$ arises by rotating each qubit in Fig.\ \ref{fig:1} by $90^\circ$ and is discussed in Ref.\ \cite{Supplement}). Estimates based on the typical dimensions of a cQED system yield $J/2\pi \sim 100\text{MHz}$. Interdigitated capacitors between the qubits might significantly increase $J$. In general, tuning $\Omega$ will also affect $J$ in a way that depends on the tuning mechanism and on the fundamental qubit parameters \cite{Supplement}. Using standard technology, upon variation of the magnetic flux, $J\propto \Omega$ for transmons, whereas, for Cooper-pair boxes, $J$ is independent of $\Omega$. Resonator A facilitates the initialization and readout of qubit 1 (with standard techniques \cite{Wallraff2004}). Resonator B would allow one to measure end-to-end correlators. However, for simplicity, we consider a system with one resonator unless otherwise noted. We mention that the proposed setup should also be implementable using the novel, high-coherence 3d cQED devices \cite{Paik2011}. Superconducting flux and phase qubits \cite{Clarke2008} can also be coupled to implement $\mathcal{H}_I$ and related Hamiltonians \cite{McDermott2005,Johnson2011}. For different proposals on the implementation of and mean-field-type experiments with the TFIC in cQED, see Refs.\ \cite{Wang2007,Tian2010}, respectively. 

In our calculations \cite{Supplement}, we frequently use the spin--free-fermion mapping for $\mathcal{H}_I$ from Refs.\ \cite{Lieb1961,Pfeuty1970}. It yields $\mathcal{H}_I = \sum_k \Lambda_k (\eta_k^\dagger \eta_k -1/2)$, where $\eta_k^\dagger$ generates a fermion of energy $\Lambda_k=2J \sqrt{1+\xi^2 -2\xi \cos k}$, and $\xi=\Omega/2J$ is the normalized transverse field. The allowed values of $k$ satisfy $\sin k N= \xi \sin k (N+1)$. For $N \rightarrow \infty$, $\mathcal{H}_I$ undergoes the second order QPT at $\xi=1$ from a ferromagnetic phase ($\xi<1$) with long-range order in $\sigma_x$ to a disordered, paramagnetic phase (for details, see \cite{Lieb1961,Pfeuty1970, Sachdev1999,Supplement}).

\paragraph*{Spectrum of the system.---} An initial experiment would likely characterize the setup by measuring the transmission spectrum $S$ of the resonator as a function of probe frequency $\omega$ and qubit frequency $\Omega$. For definiteness, we now assume that $J$ is fixed and that the transverse field $\xi = \Omega/2J$ is tunable via $\Omega$, as is the case for Cooper-pair boxes. A system with standard transmons can be shown to be confined to the paramagnetic phase (with fixed $\xi >1$), but its spectrum as a function of $\omega$ and $J\propto \Omega$  otherwise displays the same features \cite{Supplement}. To calculate $S$, we first focus on the spectrum of the bare TFIC, $\tilde{\rho}(\omega)=\int \mathrm{d}t e^{i \omega t} \langle \sigma_x^1(t) \sigma_x^1(0) \rangle$. It shows at which frequencies a field coupled to $\sigma_x^1$ can excite the chain. Assuming $g/\omega_0 \ll 1$, we then approximate the chain as a linear bath, coupled to the resonator: We replace it by a set of harmonic oscillators with the spectrum $\tilde{\rho}(\omega)$ of the TFIC. This allows us to compute $S$. Our calculations are for zero temperature. Except near the QPT, where $\mathcal{H}_I$ becomes gapless, this is experimentally well justified.

\begin{figure}[b!]
	\centering
	\includegraphics[width=\columnwidth]{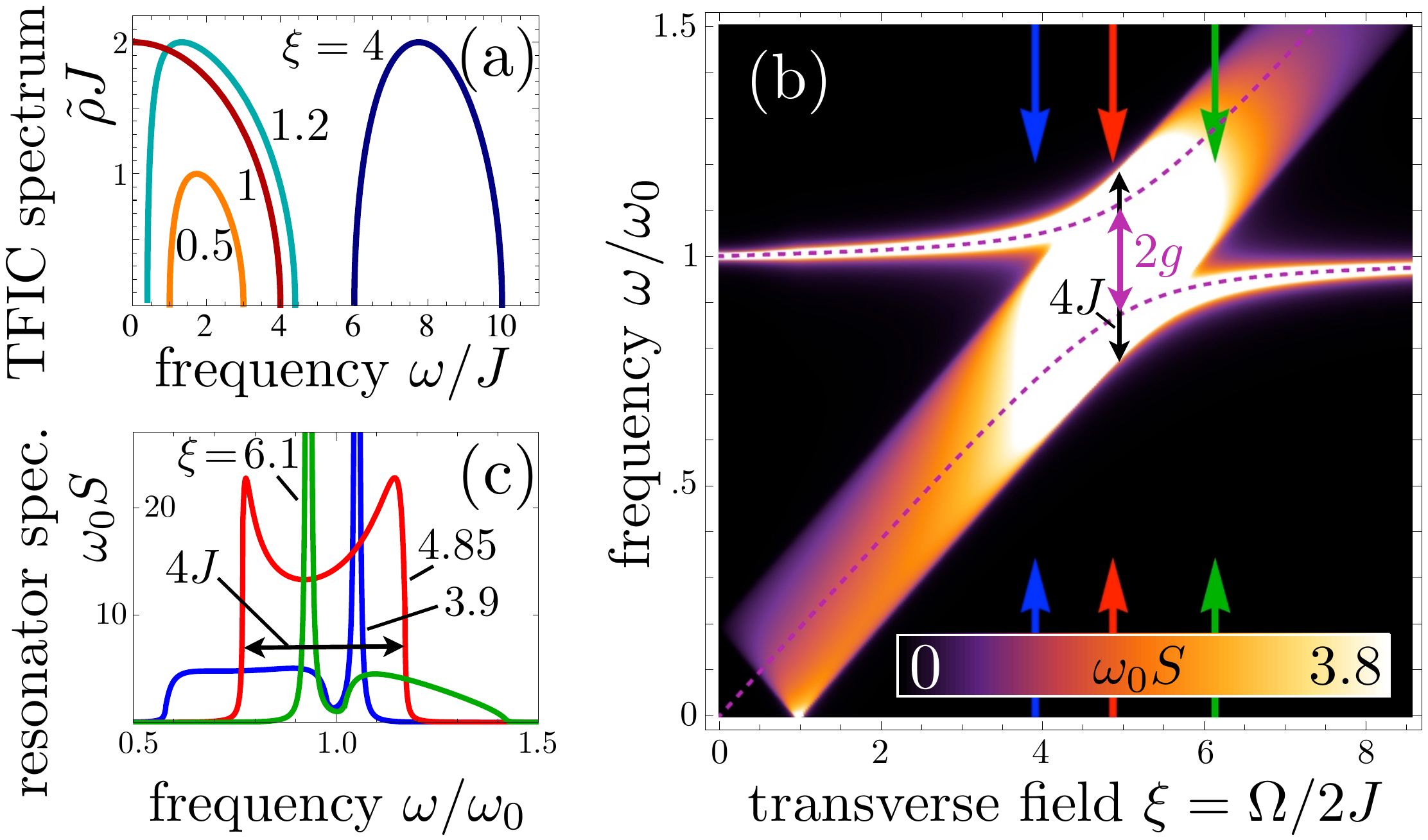} 
	\caption{
	Spectrum of the system.
	(a) Spectrum $\tilde{\rho}(\omega)= \int \mathrm{d}t  e^{i \omega t}\langle \sigma_x^1(t) \sigma_x^1(0) \rangle$
	of an isolated transverse-field Ising chain for $N\rightarrow \infty$ and normalized transverse fields $\xi= \Omega / 2 J = 4,1.2,1,0.5$. 
	(b) Spectrum $ S $ of a resonator coupled to a TFIC (as in Fig.\ \ref{fig:1}), plotted vs.\ 
	$\xi$ and $\omega$ (for $N\rightarrow \infty$). 
	The parameters used are $g=0.12$, $J=0.1$, and $ \kappa = 10^{-4}$ (in units of $\omega_0$).
	For better visibility of the features, values $> 3.8$ are plotted in white. 
	The dashed lines represent the excitation energies of $\mathcal{H}$ for $N=1$.
	(c) $ S$ vs $\omega$ for $\xi =3.9,4.85,6.1$ (blue, red, and green lines, respectively).
	These lines correspond to cuts along the arrows in (b). 
}
	\label{fig:2}
\end{figure}

For finite $N$, the calculated spectrum $\tilde{\rho} (\omega)$ would consist of discrete peaks. In an experiment, they would be broadened by decay and, for large $N$, the measured spectrum would be continuous. This can be modeled by taking $N \rightarrow \infty$ in our calculations. In that case, 
\ma{
\tilde{\rho}(\omega)&= 2\pi \delta(\omega)  \Theta (1-\xi) (1-\xi^2)  \notag\\
&+ \dfrac{4 \xi}{\omega} \mathrm{Re} \sqrt{1-\cos^2 k(\omega)} \label{eq:rhoxxomegacont}
}
for $\omega \geq 0$, and $\tilde{\rho}(\omega<0)=0$. Here, $\Theta(x)$ is the Heaviside step function, and $\cos k(\omega) = [1+\xi^2 -(\frac{\omega}{2J})^2]/2\xi$. The delta function for $\xi<1$ is due to the nonzero mean value of $\mathrm{Re}\langle \sigma_x^1(t) \sigma_x^1(0) \rangle$ in this phase. We plot $\tilde{\rho}(\omega)$ for several $\xi$ in Fig.\ \ref{fig:2}(a). For $\xi > 1$ ($\xi<1$), $\tilde{\rho}$ has a width of $4J$ ($4J\xi$), the bandwidth of the $\Lambda_k$. This might be helpful to measure $J$. At $\xi=1$, $\tilde{\rho}$ becomes gapless and, thus, carries a clear signature of the QPT. The loss of normalization for $\xi=0.5$ is compensated by the delta function in \eqref{eq:rhoxxomegacont}. This is required by a sum rule for $\tilde{\rho}$ and can be understood: In the ordered phase, the ground state $\ket{0}$ of the TFIC becomes similar to a $\sigma_x$ eigenstate. Thus, driving via $\sigma_x^1$ is less efficient in causing excitations out of $\ket{0}$, but a static force on $\sigma_x^1$ will change the energy of $\ket{0}$. We note that, for all $\xi$, $\tilde{\rho}(\omega)$ has its maximum where the band $\Lambda_k$ has zero curvature (and maximum slope). Thus, most $\eta_k$ excitations of the TFIC have a nearly uniform velocity $v_0=\max [\mathrm{d}\Lambda_k/\mathrm{d}k]$ ($v_0=2 J \xi$ for $\xi<1$ and $v_0=2 J $ for $\xi>1$), which will be important below.

We obtain resonator spectrum $S(\omega)$ in terms of $\tilde{\rho}(\omega)$, 
\ma{
S (\omega) =  \dfrac{4 \Theta (\omega)  [\kappa + g^2  \tilde{\rho}(\omega)  ]}{[\omega^2/\omega_0 - \omega_0 - 4g^2 \chi(\omega^2)]^2 + [\kappa  + g^2  \tilde{\rho}(\omega) ]^2}. \label{eq:Scoupled}
}
Here, $\chi(\omega^2)$ denotes the principal value integral $\chi(\omega^2) = 1/(2\pi) \int \mathrm{d} \Omega \tilde{\rho}(\Omega) \Omega/(\omega^2-\Omega^2)$ and $\kappa $ is the full linewidth at half maximum of the Lorentzian spectrum of the uncoupled ($g=0$) resonator. Our calculation uses tools that are explained, e.g., in \cite{Marquardt2005}. It actually also applies when the resonator couples to a different system, with another spectrum $\tilde{\rho}(\omega)$. We plot $S$ as function of $\omega$ and $\xi$ in Fig.\ \ref{fig:2}(b). For comparison, we also plot the resonances of the Jaynes-Cummings model, as they have been observed in numerous cQED experiments (dashed lines; case $N=1$ in $\mathcal{H}$). As long as the spectrum $\tilde{\rho}(\omega)$ of the chain does not overlap the resonator frequency $\omega_0$, there is a dispersive shift analogous to the off-resonant single-qubit case. Here, the chain causes only a small but broad side maximum and hardly modifies the dominant Lorentzian [green and blue lines in Fig.\ \ref{fig:2}(c)]. If the chain comes into resonance, this changes dramatically, and $S(\omega)$ takes on large values over a region of width $\sim 4J$. For our choice of parameters, $S (\omega)$ develops a slightly asymmetric double peak structure [red line in Fig.\ \ref{fig:2}(c)]. This is again reminiscent of the Jaynes-Cummings doublet, but now the peaks are split by $4J$ rather than $2g$. We emphasize that the shape of the spectrum on resonance depends significantly on the ratio $g/J$. The larger $g/J>1$, the closer the system resembles the single-qubit case (corresponds to $J=0$). If $g/J<1$, the double-peak vanishes and one observes a Lorentzian around $\omega_0$ with width $2g^2/J$ (for $g^2/J \gg \kappa$). This is because the resonator irreversibly decays into the chain, whose inverse bandwidth $\propto 1/J$ sets the density of states at $\omega \approx \omega_0$ and so determines the decay rate (for plots on both limiting cases and for finite $N$, see Ref.\ \cite{Supplement}).

\paragraph*{Propagation of a localized excitation.---} 
Off resonance, chain and resonator are essentially decoupled. In this situation, our setup allows one to study nonequilibrium dynamics in the TFIC. The resonator can be used to dispersively read out the first qubit. For measurements, this qubit must be detuned (faster than $2 \pi/J$) from the chain so that it dominates the dispersive shift of the resonator \cite{Wallraff2004} and decouples from the chain's dynamics.

\begin{figure}[b!]
	\centering
	\includegraphics[width=\columnwidth]{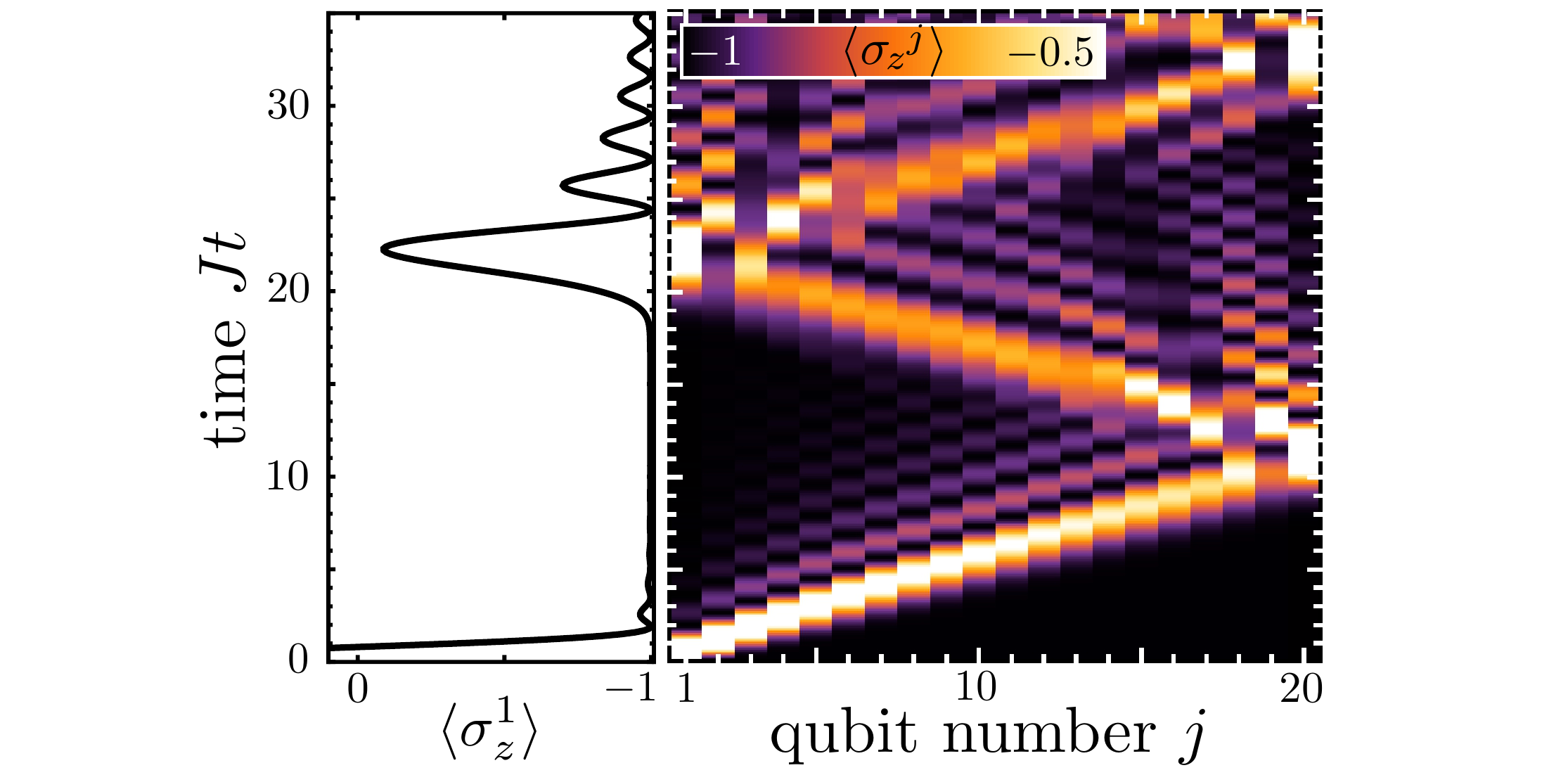}
	\caption{
	Propagation of a localized excitation.
	Right: nonequilibrium time evolution of $\langle \sigma_z^j \rangle$ for all qubits $j$ of a transverse Ising chain of length 
	$N=20$ in a normalized transverse field $\xi= \Omega/2J = 8$ (paramagnetic phase) after the first spin has been flipped. 
	Values $>-0.5$ are plotted in white. 
	Left: separate plot of $\langle \sigma_z^1 \rangle$ on the same time scale. 
	This quantity can be measured in the setup of Fig.\ \ref{fig:1}.}
	\label{fig:3}
\end{figure}

First, we focus on the nonequilibrium dynamics of the chain after a local excitation has been created. As the resonator couples only to one qubit, the initialization of the system is easy. We assume that the chain is far in the paramagnetic phase ($\xi \gg 1$). Hence, $\langle \sigma_z^j \rangle \approx -1$ in its ground state. By applying a fast ($\sim 1\text{ns}$) $\pi$ pulse, the first spin of the chain can be flipped without affecting the state of the other qubits (if $J/2\pi \ll 1\text{GHz}/2\pi$, or if the first qubit is detuned from the others for initialization). We model the state of the system immediately after the $\pi$ pulse by $\sigma_x^1 \ket{0}$, where $\ket{0}$ is the ground state of the TFIC. The time evolution of the qubit excitations $\langle \sigma_z^j \rangle$,
\ma{
\langle \sigma_z^j \rangle (t) = \bra{0} \sigma_x^1 e^{i \mathcal{H}_I t} \sigma_z^j e^{-i \mathcal{H}_I t} \sigma_x^1 \ket{0},
\label{eq:Observable_Spinwaves}
} 
is plotted in Fig.\ \ref{fig:3} for a chain with $N=20$ and $\xi=8$ (right panel). The experimentally measurable trace of $\langle \sigma_z^1 \rangle(t)$ is singled out on the left hand side. Due to the qubit-qubit coupling, the excitation propagates through the chain, is reflected at its end, and leads to a distinct revival of $\langle \sigma_z^1 \rangle$ at $J t_R \approx N$. Assuming $J/2 \pi=50$ MHz, we find $t_R \approx 64 $~ns for $N=20$, which is safely below transmon coherence times. Note that the excitation propagates with velocity $v_0 = 2 J$. This is because it consists of many excitations in $k$ space, and most of them have velocity $v_0$.

\paragraph*{Quench dynamics.---}
An appealing application of our system would be to observe its nonequilibrium dynamics after a sudden change of the transverse field $\xi = \Omega/2 J$. By using fast flux lines, changes of $\Omega$ have been achieved practically instantaneously on the dynamical time scale of a cQED system (without changing the wave function) \cite{DiCarlo2010,Fedorov2012,Reed2012}. In our setup, such a change amounts to a (global) quantum quench of $\xi$ if $J \not\propto \Omega$. This condition can be fulfilled by using qubits whose Josephson and charging energies \cite{Clarke2008} have a ratio $E_J/E_C \lesssim 10$ \cite{Supplement}, that is, Cooper-pair boxes or transmons slightly out of their optimal parameter ratio \cite{Koch2007}. In this regime, the tuning of $J$ with $\Omega$ is weak (vanishes for Cooper-pair boxes). Since it would only lead to a rescaling of time by a factor $\sim 1$, we assume in the following that $J$ is independent of $\Omega$ and consider quantum quenches of $\xi$ in our system. Quantum quenches in the TFIC have been studied theoretically, e.g. in \cite{Barouch1970,Igloi2000,Calabrese2006_2007,Igloi2011}. One usually assumes that for $t<0$ the system is in the ground state $\ket{0}_a$ of the Hamiltonian $\mathcal{H}_{I,a}$ at some initial value $\xi_a = \Omega_a/2J $. At $t=0$, $\xi$ is changed to $\xi_b= \Omega_b/2J $, and the time evolution under the action of $\mathcal{H}_{I,b}$ is investigated.

\begin{figure}[b!]
	\centering
	\includegraphics[width=\columnwidth]{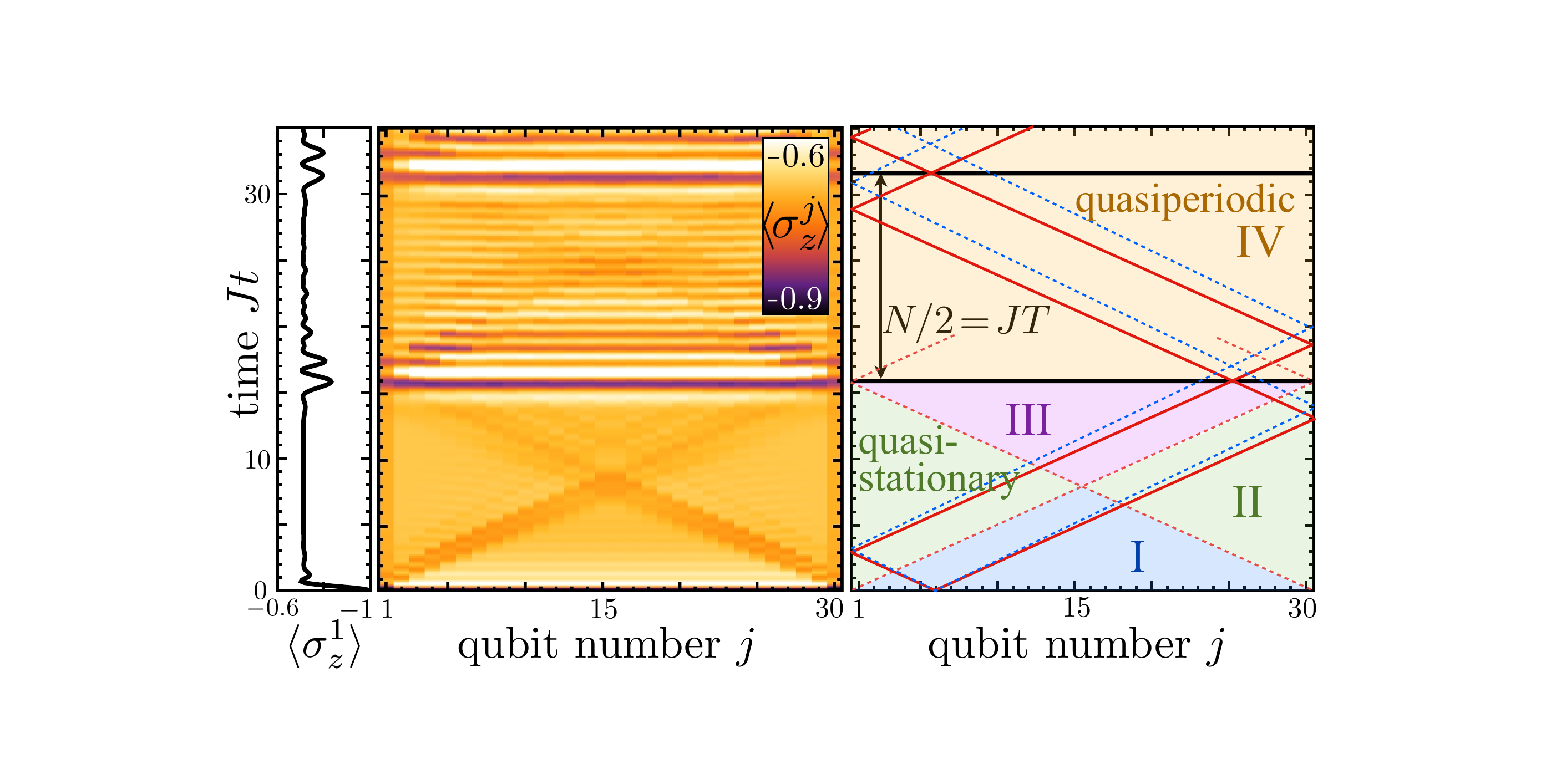}
	\caption{ 
	Behavior after a quench:
	time evolution of the magnetization $\langle \sigma_z^j \rangle$ in a TFIC of length $N=30$ 
	after a quench of the normalized transverse field $\xi = 8 \rightarrow 1.2$ (center) with a schematic plot (right) and the 
	measurable observable $\langle \sigma_z^1 \rangle$ singled out (left) on the same time scale. 
	Values $<-0.9$ ($>-0.6$) are plotted in black (white).
	}
	\label{fig:4}
\end{figure}

In the following, we focus on the dynamics of the experimentally easily accessible observable $\langle \sigma_z^1 \rangle$ after quenches within the paramagnetic phase. This corresponds to our estimates for realistic values of $J$. The main difference of quenches involving the ferromagnetic phase would be a modified dynamical time scale due to the different value of $v_0$. Figure \ref{fig:4} shows the magnetization $\langle \sigma_z^j \rangle (t)$ after quenching $\xi$ (center). In region I (see schematic plot, right), the magnetization first increases and then oscillates with decreasing amplitude. Here, it is virtually identical with the overall magnetization of a cyclic TFIC with $N\rightarrow \infty$ calculated in Ref.\ \cite{Barouch1970}, and would, for $N\rightarrow \infty$, approach a constant value. This is in line with predictions from conformal field theory \cite{Calabrese2006_2007}. However, at $t=j/v_0 $ and $t=(N-j)/v_0$ (dashed red lines in the schematic plot), where $v_0= 2J$ as before, the magnetization has dips. They are followed (in regions II and III) by a relaxation similar as in region I to the same asymptotic value (see Ref.\ \cite{Supplement} for a zoomed-in plot). Near the system boundaries, the magnetization reaches and stays at this value for a considerable time before undergoing a revival. A sharp oscillation across the entire chain at  $T=N/v_0$ subsequently decays. Revivals reoccur (quasi-)periodically with period $T$ (region IV), but this behavior is smeared out for large times (not plotted). These phenomena are reflected in the measurable observable $\langle \sigma_z^1 \rangle(t)$ (left panel) and take place on a time scale of $\sim 0.1$~$\mu$s for $N=30$ and $J/2 \pi=50$ MHz.

Our results can be qualitatively understood in a simplifying quasiparticle (QP) picture that has already been used to calculate or interpret the (quench) dynamics of different quantities in the TFIC \cite{Sachdev1997,Igloi2000,Calabrese2006_2007,Igloi2011}. In the paramagnetic phase, the QPs correspond to spins pointing in the $+z$ direction. They are created in pairs by the quench and ballistically move with velocities $\pm v_0$ with reflections at the boundaries. Further, only contiguously generated QPs are correlated. After an initial transient, any given site will be visited only by uncorrelated QPs, originating from distant places. This leads to the relaxation of the magnetization to a steady-state value in region I that would be characterized by a certain static density of uncorrelated QPs. However, once correlated QPs meet again due to reflections at the boundaries, coherences are recreated and show up in oscillation revivals. This happens, first, at multiples of $T$ (black solid lines in the schematic plot) when all QP trajectories cross their momentum-inverted counterparts (the solid red lines show an example) and, second, along the trajectories of QP pairs generated at the boundaries. Such QPs travel together as one partner is reflected at $t \approx 0$ (dashed red lines; not plotted in region IV for clarity). The periodicity of the trajectories should lead to periodic revivals for $t>T$. This is indeed observed approximately, although finally the velocity dispersion of the QPs renders the time evolution quasiperiodic. Finally, QP trajectories cannot intersect at $j=1,N$. The density of (incoherent) QPs is thus lower here than for bulk sites, yielding an appreciably lower quasistationary value.

\paragraph*{Discussion and outlook.---} The setup and the experiments we have proposed might help to establish the simulation of interacting quantum many-body systems as a new paradigm in circuit QED and to bring parts of the theoretical discourse in nonequilibrium physics closer to observation. The phenomena discussed here are based on realizable system parameters and should occur within the system's coherence time. Given the readout capabilities in cQED (e.g.\ \cite{Vijay2011}), their measurement should be feasible, for instance, because single-shot readout is not required. Once an actual implementation sets some boundary conditions, the choice of system parameters can be further optimized. We have numerically verified that all presented results are robust against disorder up to a few percent in $\Omega$ and $J$ \cite{Viehmann2012ip}. Detuning individual qubits, however, would allow one to create arbitrary potentials for the excitations, study the interplay of Anderson localization and many-body physics, or change the effective chain length. Using a second resonator, the dynamics of the end-to-end correlator $\langle \sigma_x^1 \sigma_x^N \rangle $ (indicating long-range order) could be measured (see Ref.\ \cite{Supplement}). Many other experiments are conceivable with our setup, such as suddenly coupling two isolated chains (and other local quenches) or even parameter ramps through the QPT, with Kibble-Zurek defect creation. We note also that hitherto unexplored measurement physics could be studied when the first qubit is not detuned from the chain, like resolving many-body eigenstates or the quantum Zeno effect in a many-body system. Once the setup is properly understood, it will be easy to break the integrability of our model in a controlled way (e.g., via longer-range couplings). This would push our cQED quantum simulator into a regime beyond classical computational capabilities, where further open questions about nonequilibrium dynamics can be addressed, such as thermalization and diffusive transport. Furthermore, going to 2d or 3d introduces new design options, for instance, frustrated lattices.

We thank I.\ Siddiqi, R.\ Vijay, A.\ Schmidt, and N.\ Henry for discussions. O.V.\ thanks the QNL group at UC Berkeley for their hospitality. Support by NIM, the Emmy-Noether Program, and the SFB 631 of the DFG is gratefully acknowledged.

\clearpage
\stepcounter{myequation}
\stepcounter{myequation2}

\renewcommand{\theequation}{S.\arabic{equation}}
\renewcommand{\thefigure}{S\arabic{figure}}

\thispagestyle{empty}
\onecolumngrid
\begin{center}
{\fontsize{12}{12}\selectfont \textbf{Supplementary Material for "Observing the nonequilibrium dynamics of the quantum transverse-field Ising chain in circuit QED"\\[5mm]}}
{\normalsize Oliver Viehmann,$^1$ Jan von Delft,$^1$ Florian Marquardt$^2$\\[1mm]}
{\fontsize{9}{9}\selectfont
\textit{	$^1$Physics Department,             
			 Arnold Sommerfeld Center for Theoretical Physics,
             and Center for NanoScience,\\
             Ludwig-Maximilians-Universit\"at,
             Theresienstra{\ss}e 37,
             80333 M\"unchen, Germany \\
 			$^2$Institut for Theoretical Physics, Universit\"at Erlangen-N\"urnberg, Staudtstra\ss e 7, 91058 Erlangen, Germany}}
\vspace*{6mm}
\end{center}
\normalsize
\twocolumngrid

\section{Contents}
\noindent
I. The qubit-qubit coupling Hamiltonian \hfill 6 \\[8pt]
\noindent
II. Diagonalization and spectrum of the transverse-field \hfill \\  Ising chain \hfill 9 \\[8pt]
\noindent
III. Spectrum of the resonator \hfill 11 \\[8pt]
\noindent
IV. Propagation of a localized excitation in the Ising \hfill \\ \noindent chain \hfill	14 \\[8pt]
\noindent
V. Quench dynamics of the magnetization and the \hfill \\ \noindent end-to-end correlations	\hfill 15 \\[8pt]
\noindent
References	\hfill 15

\section{I. The qubit-qubit coupling Hamiltonian} \label{sec:QQcoupling}

In this section, our goal is to derive the Hamiltonian of a chain of
capacitively coupled charge-based artificial atoms as in Fig.\ 1 of the main
text from circuit theory. Both for Cooper-pair boxes (CPBs) and for transmons
(for reviews on superconducting artificial atoms, see \cite{Supp_Makhlin2001,Supp_Clarke2008}), this
Hamiltonian takes on the form of $\mathcal{H}_I$ [Eq.\ (2) of the main text]. 
Our derivation of the Hamiltonian on the basis of circuit theory enables us to analyze the dependence of $\Omega$ and $J$
(and thus of $\xi = \Omega/2J$) on the fundamental, engineerable parameters of the
artificial atoms and on an externally applied, in-situ tunable magnetic flux.

\begin{figure}[b!]
\centering
\includegraphics[width=\columnwidth]{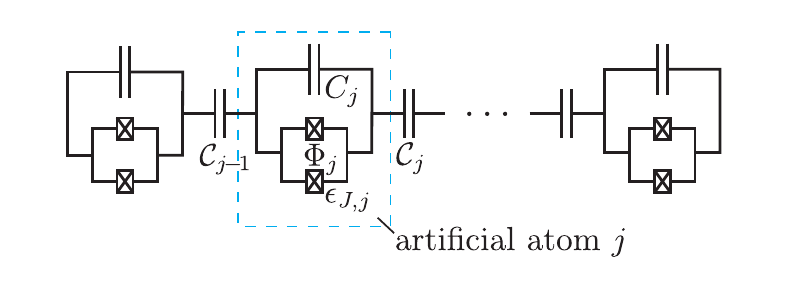}
\caption{Circuit diagram of a chain of capacitively coupled charge-based artificial
atoms as in Fig.\ 1 of the main text.}
\label{fig:S1}
\end{figure}

We model the chain of artificial atoms in Fig.\ 1 of the
main text by the circuit diagram of Fig.\ \ref{fig:S1}. The SQUID-like loop
of the $j$th artificial atom can be threaded by a (classical) external magnetic
flux bias $\Phi_j$. Its identical Josephson junctions are characterized each by a Josephson
energy $\epsilon_{J,j}$. For simplicity, we absorb the capacitances of the Josephson
junctions into the
capacitance $C_j$ between the islands of the artificial atom (which shunts the
SQUID loop). Moreover, we take into account only coupling capacitances $\mathcal{C}_{j}$ between the right island of the $j$th
artificial atom and the left island of the $j+1$st artificial atom. The mediated 
capacitive coupling between the artificial atoms corresponds to the electrostatic
coupling of the electric dipole operators of charge distributions in atomic QED \cite{Supp_CohenTannoudji2004}, 
which we have employed in the main text to motivate the Hamiltonian
$\mathcal{H}_I$. In order to be able to compare our results with previous ones \cite{Supp_Pashkin2003,Supp_Dewes2012}, we
do not assume that the artificial atoms are identical for the moment. 

We begin by considering the conceptionally important case $N=2$. This case has been
already studied for CPBs \cite{Supp_Pashkin2003} and transmons \cite{Supp_Dewes2012} in similar
setups. Using the standard approach to
circuit quantization \cite{Supp_Yurke1984,Supp_Devoret1995}, one obtains  
$ H^{(2)}= \sum_{j=1}^2 (q_j^2/2 \tilde{C}_j - E_{J,j}^\Phi \cos
2e \phi_j )
+ q_1 q_2/\tilde{\mathcal{C}}$.
Here, $\hbar =1$, $\phi_j$ and $q_j$ are the conjugate quantum flux and charge 
variables, $[\phi_j, q_j] = i$, and $e$ is the elementary charge. We have defined
$\tilde{C}_j = C_{\ast}^2/(C_{j} +
\mathcal{C})$, $\tilde{\mathcal{C}} = C_\ast^2/\mathcal{C}$, and $C_{\ast}^2 =
C_1 C_2 + C_1 \mathcal{C} + C_2 \mathcal{C}$ (for $N=2$, we drop the index $1$
from $\mathcal{C}_1$ and related quantities like $\tilde{\mathcal{C}}_1$). Furthermore,
$E_{J,j}^\Phi = E_{J,j}(\Phi_{j}) = 2\epsilon_{J,j} \cos ( \pi \Phi_{j}/\Phi_0)$, where
$\Phi_0$ is the superconducting flux quantum. As usual, we introduce charging
energies $E_{C,j} = e^2/ 2 \tilde{C}_j$, number and phase operators $\hat{n}_j =-
q_j/2e$ and $\varphi_j = -2e \phi_j$ (see, e.g., \cite{Supp_Devoret1995}), and a coupling energy $E_\mathcal{C} = e^2/2
\tilde{\mathcal{C}}$. The effects of possible gate
voltages that might bias the superconducting islands of an artificial atom are taken into
account by introducing offset charges $n_{b,j}\in \mathds{R}$ (in units of $2e$) and substituting
$\hat{n}_j \rightarrow \hat{n}_{b,j} \equiv \hat{n}_j - n_{b,j}$ (possible gate capacitances are assumed
to be absorbed in $E_{C,j}$ and $E_{\mathcal{C}}$). With these substitutions, 
\ma{
H^{(2)}= h_1 + h_2   + 8 E_{\mathcal{C}} \hat{n}_{b,1}\hat{n}_{b,2}.\label{eq:Supp_H22}
}
The $h_j = 4 E_{C,j}\hat{n}_{b,j}^2 - E_{J,j}^\Phi \cos
\varphi_j $ describe the energies of two isolated artificial
atoms. The eigenfunctions (in the $\varphi_j$-basis) and eigenvalues of $h_j$ are
Mathieu's functions and characteristic values \cite{Supp_Cottet2002,Supp_Devoret2003,Supp_Koch2007}. Their numerical
values can be determined with arbitrary precision for all parameters $E_{C,j}$,
$E_{J,j}^\Phi$, and $n_{b,j}$ (and all $\varphi_j$) and are implemented in standard math programs. 
Taking the ground state $\ket{g_j}$ and the first
excited state $\ket{e_j}$ of $h_j$ to be eigenstates of $\sigma_z^j$ and restricting the
Hilbert space to these qubit bases, the Hamiltonian of the system becomes (up to a
constant)
\ma{
H^{(2)}= \sum_{j=1}^2 \dfrac{\Omega_j}{2} \sigma_z^j + 8 E_{\mathcal{C}}
\prod_{j=1}^2 \sum_{m_j,n_j} (\hat{n}_{b,j})_{m,n} \ket{m_j} \! \bra{n_j}.
\label{eq:Supp_H23}
}
Here, $\Omega_j$ is the difference between the qubit eigenenergies, $(\hat{n}_{b,j})_{m,n} =
\bra{m_j}\hat{n}_{b,j}\ket{n_j}$, and $m_j, n_j \in \{g_j, e_j\}$.
Using the explicit forms of $\langle \varphi_j |
m_j \rangle$ from \cite{Supp_Cottet2002,Supp_Devoret2003,Supp_Koch2007} and $\hat{n}_j = -i \partial /
\partial \varphi_j$, the $(\hat{n}_{b,j})_{m,n}$ are found to be real and can be
numerically calculated. In general, the $\hat{n}_{b,j}$ have diagonal elements
in our choice of basis. However, for the most common types of charge-based artificial atoms,
CPBs and transmons, $H^{(2)}$ takes on the form of $\mathcal{H}_{I}$ for
$N=2$, which is insightful to consider before returning to the general case.

CPBs are characterized by $4 E_{C,j} \gg E_{J,j}^\Phi$. Since we are interested only in the
low-energy sector of the Hilbert space of \eqref{eq:Supp_H22}, this condition allows us, in
good approximation, to restrict the Hilbert space to the number states $\{\ket{n_j},
\ket{(n+1)_j}\}$. Here, $n_j= \lfloor n_{b,j} \rfloor$ \cite{Supp_Makhlin2001}. Without loss of generality, one
can choose $n_{b,j} \in [0,1[$. This restriction leads from Eq.\ \eqref{eq:Supp_H22} to
\begin{align}
H_{CPB}^{(2)}&= \sum_{j=1}^2 \Big[ 4 E_{C,j} ( n_{b,j}^2 \ket{0_j}\! \bra{0_j} + (1-
n_{b,j})^2 \ket{1_j} \!\bra{1_j})  \nonumber \\
& \hspace{10pt} - \frac{E_{J,j}^\Phi}{2} (\ket{0_j}\! \bra{1_j} +
\mathrm{H.c.}) \Big]\nonumber \\ 
&+ 8 E_{\mathcal{C}} \prod_{j=1}^2  \sum_{n_j=0}^1 (n_j - n_{b,j}) \ket{n_j} \!
\bra{n_j} ,
\end{align}
in close similarity to the Hamiltonian derived in \cite{Supp_Pashkin2003}. If 
the CPBs are operated as usual at the charge degeneracy points $n_{b,j} = 1/2$
(to decrease charge noise), $\ket{g_j}$ ($\ket{e_j}$) is an (anti-)symmetric
superposition of $\ket{0_j}$ and $\ket{1_j}$. We drop constants, identify
$\ket{0_j} \! \bra{1_j} = \sigma_-^j$, and rotate the coordinate system
by $\pi/2$ around the $y$-axis (clockwise). This brings $H_{CPB}^{(2)}$ into the form of
\mbox{Eq.\ \eqref{eq:Supp_H23}}, 
\ma{
H_{CPB}^{(2)} = \sum_{j=1}^2 \frac{ E_{J,j}^\Phi}{2}\sigma_z^j + 2 E_\mathcal{C} \sigma_x^1
\sigma_x^2.
}
This  Hamiltonian also has the form of $\mathcal{H}_I$ for $N=2$ (since $\hat{n}_{b,j} $
transforms into $\sigma_x^j/2$ under the present assumptions). That is, in the case of CPBs,
the transition frequencies $\Omega_j$ are simply given by
$E_{J,j}^\Phi$ and flux-tunable. The qubit-qubit coupling $J=2E_\mathcal{C}$
depends only on the capacitances of the system and is independent of $\Phi_j$ and the
qubit transition frequencies ($\mathrm{d} J/ \mathrm{d} \Omega_j = 0$). Thus, the
normalized transverse field $\xi = \Omega/2 J$ (for $\Omega_1 = \Omega_2$) is strictly
linear in $\Omega$.

For transmons-qubits \cite{Supp_Koch2007}, which are characterized by $4E_{C,j} \ll
E_{J,j}^\Phi$, (i) expanding the $\cos \phi_j$ terms in $h_j$ of Eq.\ \eqref{eq:Supp_H22}
and (ii) dropping the boundary condition $\psi_{j,m} (\varphi_j) = \psi_{j,m}
(\varphi_j + 2\pi)$ on the eigenfunctions of $h_j$ provides a good approximation \cite{Supp_Koch2007}.
Note that due to (ii), the effect of the offset charges $n_{b,j}$ is completely
suppressed since the $\hat{n}_{j}$ and the biased number operators $\hat{n}_{b,j}$ are
equivalent canonical variables, $[\varphi_j, \hat{n}_{b,j}] = [\varphi_j,
\hat{n}_{j}]= i$. This is justified as the dependence of the qubit properties on the offset charges is exponentially
suppressed with increasing ratio $E_{J,j}^\Phi/E_{C,j}$ \cite{Supp_Koch2007} (in
reality, gate voltages do not have to be applied to transmons). Thus, we now aim to derive the parameters $\Omega_j$ and $(\hat{n}_{b,j})_{m,n}$ occuring in Eq.\ \eqref{eq:Supp_H23} from
Eq.\ \eqref{eq:Supp_H22} with
\ma{
h &\approx 4 E_{C} \hat{n}_{b}^2 - E_{J}^\Phi (1- \varphi^2/2! + \varphi^4/4!)
\nonumber \\
&= \Omega_0 (a^\dagger a + 1/2) - \alpha \Omega_0 (a^\dagger + a)^4/4! +
\mathrm{const.}, \label{eq:Supp_HTexpand}
}
in a perturbation expansion in $\alpha= (E_{C}/ 2 E_{J}^\Phi)^{1/2} \ll 1$. Here and in
the following, we drop the index $j$ where not essential. Note that
$\alpha$ is
proportional to an approximate expression for a transmon's `relative anharmonicity'
\cite{Supp_Koch2007}. We have defined
$\Omega_0=(8 E_J^\Phi E_C)^{1/2}$, $\varphi =
\sqrt{2 \alpha} (a^\dagger + a) $, $\hat{n}_b = i/\sqrt{8 \alpha} (a^\dagger -
a)$, and $[\phi,\hat{n}_b] = i $ requires $a$ to be bosonic. This approach has been
used in \cite{Supp_Koch2007} to study a single transmon and its coupling to a microwave
resonator. To first order in
$\alpha$, $\ket{g_\alpha} = \ket{0} + \alpha/4! (3\sqrt{2} \ket{2} + \sqrt{3/2} \ket{4})$
and $\ket{e_\alpha} = \ket{1} + \alpha/4! (5\sqrt{6} \ket{3} + \sqrt{15/2} \ket{5})
$, where $\ket{m}$ is now an eigenstate of $a^\dagger a$. We
substitute Eq.\ \eqref{eq:Supp_HTexpand} and the above expression for $\hat{n}_b$ into 
Eq.\ \eqref{eq:Supp_H22} and expand the resulting transmon-approximation $H_t^{(2)}$ of
$H^{(2)}$ in the qubit basis spanned by $\ket{g_\alpha}$ and $\ket{e_\alpha}$. 
Dropping constants and all terms $\propto \alpha^x$ with $x>1$,
and rotating the coordinate system counter-clockwise by $\pi/2$
around the $z$-axis leads to
\ma{
H_t^{(2)} = \sum_{j=1}^2 \dfrac{\Omega_{0,j} (1- \alpha_j/2)}{2} \sigma_z^j +
E_\mathcal{C} \prod_{j=1}^2 \dfrac{(1- \alpha_j/4)}{\sqrt{\alpha_j}}
\sigma_x^j.\nonumber 
} 
This transmon approximation of Eq.\ \eqref{eq:Supp_H23} also has the form of
$\mathcal{H}_I$ for $N=2$. We remark that in $0$th order perturbation theory, where the transmons are harmonic
oscillators, the terms in parentheses in $H_t^{(2)}$ are equal to $1$.
However, the term $1/\sqrt{\alpha_1 \alpha_2}$, stemming from the product of the $\hat{n}_{b,j}$
operators, is present. The $0$th order result corresponds to the Hamiltonian derived
in \cite{Supp_Dewes2012} for their system of coupled transmons. To first order in $\alpha$,
the transmon transition frequencies are given by $\Omega_j = \Omega_{0,j}(1-
\alpha_j/2) = (8E_{J,j}^\Phi E_{C,j})^{1/2} - E_C$ \cite{Supp_Koch2007}. They are
flux-tunable via $(E_{J,j}^\Phi)^{1/2}$
(rather than $\Omega_j \propto E_{J,j}^\Phi$ as for CPBs).
For transmons, the qubit-qubit coupling is given by
$J= E_\mathcal{C} \prod (1- \alpha_j/4)/\sqrt{\alpha_j}$. Importantly, this
$J$ depends also on external fluxes via
$\alpha_j \propto (E_{J,j}^\Phi)^{-1/2}$ (and on the
transition frequencies via $\alpha_j = 2E_{C,j}/\Omega_{0,j}$).
Since the physical properties of a uniform TFIC are essentially determined by the
normalized transverse field $\xi = \Omega/2J$  (the absolute values of $\Omega$ and $J$ only set the
dynamical time scales), we use our perturbative results to study the tunability of
$\xi $ for identical transmons. We insert our first-order results for $\Omega$ and
$J$ into $\xi$ and expand
$\xi \approx (\Omega_0/2 E_\mathcal{C}) [\alpha -  \alpha^3/16 +\mathcal{O}
(\alpha^4)]$, where we have set $E_{C(J),1}=E_{C(J),2}$.
The overall factor $\alpha$ comes
from the nominator of $J$ and is not due to the nonlinear perturbation of the system
as argued above. Factoring out $\alpha = 2E_C/\Omega_0$ yields 
\ma{
\xi \approx \dfrac{E_C}{E_\mathcal{C}} (1 - \alpha^2/16 + \mathcal{O} (\alpha^3))
\approx \dfrac{E_C}{E_\mathcal{C}}.\label{eq:Supp_xiappr} 
}
That is, the first order corrections to $\Omega$ and $J$ in $\alpha$ exactly cancel. 
For transmons, flux-tunability of $\xi$
is a second-order effect, via $\alpha^2 = E_C/2E_J^\Phi$. To roughly estimate the strength of
this effect, we consider the contribution of the first-order approximations of
$\Omega$ and $J$ to it. Note that the second-order
approximations of $\Omega$ and $J$ actually also contribute to the leading flux-dependent term ($\propto
\alpha^2$) of $\xi$. If one requires the transmons to remain in their optimal
working regime $20 \lesssim E_J^\Phi/E_C $ \cite{Supp_Koch2007}, this contribution leads to a
tunability $\Delta \xi/\xi \approx \alpha^2/(16-\alpha^2) < 0.2 \%$. 
Thus, one may expect that strongly tuning 
$\xi$ by changing the flux bias will require to leave the optimal transmon working
regime, and possibly even to go beyond the
validity regime of Eq.\ \eqref{eq:Supp_HTexpand}. Therefore, we now come back to the
general case of Eq.\ \eqref{eq:Supp_H22}. 
Before doing so, we remark that $\xi  \approx E_C/E_\mathcal{C} =
(C+\mathcal{C})/\mathcal{C} >1$. This indicates that the ferromagnetic phase ($\xi<1$)
cannot be reached with transmons.

\begin{figure}
\centering
\includegraphics[width=\columnwidth]{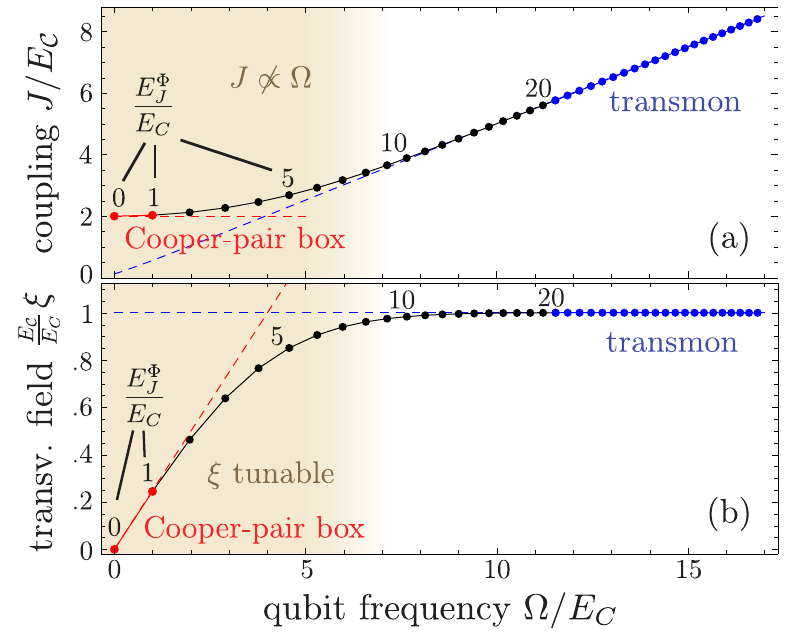}
\caption{(a) Qubit-qubit coupling $J$ and (b) normalized transverse field $\xi = \Omega/2J$
vs.\ qubit transition frequency $\Omega$ for two identical charge qubits operated
at the charge degeneracy point. The system is characterized by the charging
energy $E_C$ and the flux-tunable total Josephson energy $E_J^\Phi$ of a qubit, and
by the capacitive coupling energy $E_\mathcal{C}$. The ratio $E_J^\Phi/E_C$ fully
determines a point on each axis (i.e., the quantities $\Omega/E_C$, $J/E_\mathcal{C}$, and $\frac{E_\mathcal{C}}{E_C} \xi$). The dots correspond to the integer values
$0,1,\ldots,40$ of $E_J^\Phi/E_C$. The solid lines are a guide to the eye. 
Dashed, approximate analytical results for the limits $E_J^\Phi /4E_C \ll 1$
(Cooper-pair boxes) and $E_J^\Phi/4 E_C \gg 1$ (transmons).\vspace{-0.2cm} }
\label{fig:S2}
\end{figure}

It turns out that at the charge degeneracy point $n_{b}= 1/2$, the biased charge
operator $\hat{n}_{b} = \hat{n} - n_{b}$ has only off-diagonal
elements in the basis chosen in Eq.\ \eqref{eq:Supp_H23}. Consequently, $H^{(2)}$ has the form of
$\mathcal{H}_I$ (at $N=2$) for all ratios $E_{J,j}^\Phi/E_{C,j}$. This enables us to
interpolate between the charge-degenerate CPB case and the transmon case (where the $n_{b,j}$
become irrelevant): Assuming identical qubits, we vary the ratio
$E_{J}^\Phi/E_C$ at $n_b = 1/2$. We numerically calculate $J = 8 E_\mathcal{C}
[(\hat{n}_{b})_{g,e}]^2$ and $\Omega$ as functions of $E_J^\Phi/E_C$. Then we plot
$J$ vs.\
$\Omega$ [Fig. \ref{fig:S2}(a)] and $\xi$ vs.\ $\Omega$ [Fig.\ \ref{fig:S2}(b)].
Additionally, we plot the approximate results that we have gained analytically for CPBs
and transmons. To obtain $J$ as a function of $\Omega$ from our analytical results
for transmons, we employ our approximation for $J$ to first order in $\alpha$. In this
approximation, we replace $\alpha \approx 2
E_C/(\Omega + E_C)$, making use of the first order approximation for $\Omega$.
The plots show that for $E_J^\Phi/ E_C \gtrsim 10$, the qubit-qubit coupling $J$
becomes proportional to $\Omega$, and the normalized transverse field $\xi$ ceases to
be flux-tunable. For quenching $\xi$ by changing the flux bias one therefore
has to engineer $E_J^\Phi/E_C \lesssim 10$. In this regime, the artificial atoms
start to loose their insensitivity to charge noise, which is a distinguishing property of transmons. 
For instance, at $E_J^\Phi/E_C = 10$, 
$[\max \Omega(n_b) - \min \Omega(n_b) ]/ \overline{\Omega(n_b)} \approx 3\%$. Here,
$\overline{\Omega(n_b)}$ denotes the mean qubit transition frequency, averaged over
all possible bias charges $n_b$. 
However, the characteristic features of the quench dynamics of our circuit QED
quantum simulator occur on short timescales (see main text), so that one should get along with
the reduced dephasing times of charge qubits in this regime (compared to usual
transmons). For example, an energy
relaxation time $T_1$ of $ \sim 7 \mu \rm{s}$ and a dephasing time $T_2$ of $ \sim 500
\rm{ns}$ have been reported even for a CPB (at the charge degeneracy point) \cite{Supp_Wallraff2005}.
We remark that, depending on the charge bias $n_b$, $\Omega$
can be equal to the energy difference between second and first excited state of the
artificial atom, $E_{2,1}$, which would invalidate the two-level approximation for the
artificial atoms. For instance, if $n_{b} = 0.5$ as considered here, this happens at
$E_J^\Phi/E_C \approx 9.03$
\cite{Supp_Koch2007}. However, the difference of these transitions crosses
zero very steeply as a function of $E_J^\Phi/E_C$ \cite{Supp_Koch2007}. Thus, the two-level approximation
for the artificial atoms is justified as long as start or end point of the quench are
not too close to this value. We finally remark that
working with tunable coupling capacitances \cite{Supp_Averin2003} might provide an alternative to working with transmons out of their optimal
parameter range. This would allow one to tune $\xi$ via tuning $E_\mathcal{C}$.

Coming now to the general case of a chain of artificial atoms of arbitrary length, it
turns out that we can directly apply our results for $N=2$. Having written the Lagrangian of such a system
in terms of the classical variables $\phi_j$ and $\dot{\phi_j}$
\cite{Supp_Yurke1984,Supp_Devoret1995}, one finds that the canonical charge
variables $q_j$ are given by $\mathbf{q} = \underline{C} \dot{\boldsymbol\phi}$.
Here, we have defined $\mathbf{q} = (q_1,\ldots,q_N)^T$, $\dot{\boldsymbol\phi} =
(\dot{\phi}_1,\ldots,\dot{\phi}_M)^T$, and
\begin{align}
\underline{C} = 
\left(
   \begin{array}{cccccc}
     C+ \mathcal{C} & -\mathcal{C} & 0 &  & \cdots & 0\\
     -\mathcal{C} & C + 2\mathcal{C} & -\mathcal{C} & & & \\
     0 & -\mathcal{C} & C + 2\mathcal{C} & -\mathcal{C} & & \\
     \vdots & & \ddots & \ddots & \ddots & \\ 
      & & & -\mathcal{C}&C + 2 \mathcal{C} &-\mathcal{C}\\
	0 & & & & -\mathcal{C} & C+\mathcal{C}   
   \end{array} \nonumber
\right),
\end{align}
and we have assumed that the artificial atoms are identical, $C_j = C$ and
$\mathcal{C}_j = \mathcal{C}$. Inverting $\underline{C}$ yields 
$\dot{\boldsymbol\phi}(\mathbf{q})$. With that, one obtains the Hamiltonian $H$ of the
system, which is then quantized as usual \cite{Supp_Yurke1984,Supp_Devoret1995}. To first order in
$\mathcal{C}/C$, 
\ma{
H &= \sum_{j=1}^N \left( \frac{q_j^2}{2C} - E_J^\Phi \cos 2e \phi_j \right) \nonumber
\\
& + \frac{\mathcal{C}}{C} \left( \frac{-q_1^2 -q_N^2 -\sum_{j=2}^{N-1} 2 q_j^2 +
\sum_{j=1}^{N-1} 2 q_j q_{j+1}}{2C}\right).\label{eq:Supp_HN}
}
The same steps as for $N=2$ now lead to a straightforward generalization of
Eq.\ \eqref{eq:Supp_H22}, where artificial atoms with Hamiltonian $h_j$ are coupled to
their nearest neighbours via $\hat{n}_{b,j} \hat{n}_{b,j+1}$ [for $N=2$,
Eq.\ \eqref{eq:Supp_HN} equals the first order expansion of $H^{(2)}$ above Eq.\
\eqref{eq:Supp_H22}]. To first order in $\mathcal{C}/C$, the only difference
for $N >2$ is that the effective charging energies of the artificial atoms in the
bulk of the chain ($j\neq1,N$) are slightly reduced compared to those at the surface
($j=1,N$). This is because the bulk artificial atoms couple to two neighbours. 
In reality, this surface inhomogeneity should be negligible
already because the capacitance of the surface artificial atoms is also increased by
their coupling to other parts of the circuit. Therefore, to first order in
$\mathcal{C}/C$, our derivation of the
Hamiltonian $\mathcal{H}_I$ of the TFIC from the circuit theory of
two artificial atoms also holds for larger
chains, only with a slightly renormalized $E_C$. The same is true for 
our corresponding deliberations on the dependence of
$\Omega$, $J$, and $\xi$ on the fundamental circuit quantities. 
We remark that taking into account terms of order
$(\mathcal{C}/C)^l$ introduces coupling terms $\propto q_j q_{j+l}$ in Eq.\
\eqref{eq:Supp_HN} (and, for $l>1$, renormalizes also the nearest neighbour
coupling energies $E_\mathcal{C}$ compared to the case $N=2$).
Hence, the integrability-breaking longer-range coupling decays exponentially
with distance $l$ in our system and is therefore neglected in this work. We finally
remark that nonperturbative numerical calculations strongly suggest that 
also the renormalized values of $E_C$ and $E_\mathcal{C}$ 
for $N>2$ do not allow one to
achieve $E_C/E_\mathcal{C} <1$. This means that the ferromagnetic phase cannot be
reached with transmons in the limit of large $E_J^\Phi/E_C$ [cf.\ Eq.\
\eqref{eq:Supp_xiappr}].

\section{II. Diagonalization and spectrum of the transverse-field Ising chain}
In this section, we diagonalize the Hamiltonian $\mathcal{H}_I$ [Eq.~(2) of the main
text] and calculate the qubit autocorrelator $\rho(t)=\langle \sigma_x^1(t)
\sigma_x^1(0) \rangle$ and the corresponding spectrum $\tilde{\rho}(\omega)=\int dt
e^{i \omega t} \rho(t)$. Our method and notation follow Ref.\ \cite{Supp_Lieb1961}.  

In the main text, we have focussed on a circuit QED system with ferromagnetic
qubit-qubit coupling $J>0$. Since setups with antiferromagnetic coupling are also
conceivable, we generalize in the remainder of these supplementary notes the
Hamiltonian of the transverse-field Ising chain to
\ma{
\mathcal{H}_I = \dfrac{\Omega}{2} \sum_{j=1}^N \sigma_z^j -\mathcal{J}
\sum_{j=1}^{N-1} \sigma_x^j \sigma_x^{j+1}, \label{eq:Supp_A_TFIC}
}
where $\mathcal{J}$ may be negative ($\Omega>0$ as before). We define
$J=|\mathcal{J}|$. Applying the Jordan-Wigner transformation $\sigma_j^+ =
c_j^\dagger \exp(i \pi \sum_{k=1}^{j-1} c_k^\dagger c_k )$ to $\mathcal{H}_I$ leads
to
\begin{align}
\mathcal{H}_I = - \dfrac{N \Omega}{2} + \Omega \sum_{j=1}^N c_j^\dagger c_j -
\mathcal{J}\! \sum_{j=1}^{N-1} [ c_j^\dagger c_{j+1}^\dagger\! +
c_j^\dagger c_{j+1}\! +\! \mathrm{H.c.}] ,\label{eq:Supp_cfermions}
\end{align}
with fermionic $c_j$. In this form, $\mathcal{H}_I$ can be diagonalized using the
method for diagonalizing quadratic fermionic Hamiltonians of the form 
\ma{
H = \sum_{i,j=1}^N [ c_i^\dagger A_{i,j} c_j + 1/2 (c_i^\dagger B_{i,j} c_j^\dagger +
\mathrm{H.c.})]  \label{eq:Supplement_quadratic_Hamiltonian}
}
of Ref.\ \cite{Supp_Lieb1961}. In our case, 
\begin{align}
A = 
\left(
   \begin{array}{cccccc}
     \Omega & -\mathcal{J} & 0 &  & \cdots & 0\\
     -\mathcal{J} & \Omega & -\mathcal{J} & & & \\
     0 & -\mathcal{J} & \Omega & -\mathcal{J} & & \\
     \vdots & & \ddots & \ddots & \ddots & \\ 
      & & & -\mathcal{J}&\Omega &-\mathcal{J}\\
	0 & & & & -\mathcal{J} &\Omega   
   \end{array}
\right), \label{eq:Supplement_MatrixA}
\end{align}
and $B$ is obtained by substituting $A_{i,i} = \Omega \rightarrow 0 $ and $A_{i+1,i}=-\mathcal{J} \rightarrow \mathcal{J}$ in $A$. $H$ is diagonalized by introducing new fermions $\eta_k = \sum_{j=1}^N g_{k,j} c_j + h_{k,j} c_j^\dagger$. The components $g_{k,j}$ and $h_{k,j}$ of the vectors $g_k$ and $h_k$ and the eigenvalues $\Lambda_k$ of $H$ are determined by defining normalized vectors $\phi_k = g_k + h_k$ and $\psi_k = g_k - h_k$ and solving the equations 
\begin{align}
\phi_k(A-B) = \Lambda_k \psi_k,\qquad
\psi_k(A+B) = \Lambda_k \phi_k. \label{eq:Supplement_relpsiphi}
\end{align}
For $\Lambda_k \neq 0$, this is most easily done by solving, e.g., 
\ma{
(A-B)(A+B) \phi_k = \Lambda_k^2 \phi_k
}
and calculating $\psi_k$ via Eqs.\ \eqref{eq:Supplement_relpsiphi}. Note that since
$A^T = A$ and $B^T = -B$, $\Lambda_k^2 \geq 0$ and the $\phi_k$ and $\psi_k$ can be
chosen real and orthogonal for different $k$, $\sum_j \phi_{k,j} \phi_{k^\prime,j} =
\sum_j \psi_{k,j} \psi_{k^\prime,j} = \delta_{k, k^\prime}$. For $A$ and $B$ as
defined above, one obtains
\ma{
\mathcal{H}_I &= \sum_k \Lambda_k (\eta_k^\dagger \eta_k -1/2),\\
\Lambda_k  &=2J \sqrt{1+\xi^2 -2\xi \cos k},\\
\phi_{k,j}&= A_k \sin k(N+1-j), \label{eq:Supplement_phi_explicit}\\  
\psi_{k,j}&= \mathrm{sign} \Big[ \dfrac{\mathcal{J} \sin k}{\sin k(N+1)} \Big] A_k
\sin k j , \label{eq:Supplement_psi_explicit} \\ 
A_k  &= 2 \big(2N+1- \sin [k(2N+1)]/\sin k \big)^{-1/2}.
}
Here, $\xi = \Omega/2 \mathcal{J}$ is the normalized transverse field, and the
possible values of $k$ are solutions of 
\ma{
\dfrac{\sin k N}{\sin k (N+1)} = \xi. \label{eq:Supplement_kvalues}
}
If $|\xi| \geq N/(N+1)$ ($|\xi| < N/(N+1)$), Eq.\ \eqref{eq:Supplement_kvalues} has
$N$ ($N-1$) real solutions $ \in [0,\pi]$. If $|\xi| < N/(N+1)$, there is also one
imaginary solution $k^\prime =i \kappa$ ($k^\prime= \pi + i \kappa$) for positive
(negative) $\xi$ with $\sinh \kappa N/ \sinh \kappa (N+1) = |\xi|$. These solutions
exhaust the eigenmodes of the system. Note that $\Lambda_{k^\prime} \rightarrow 0$ if
$|\xi|\rightarrow 0$ or $N\rightarrow \infty$. 

\begin{figure}
	\centering
	\includegraphics[width=0.75\columnwidth]{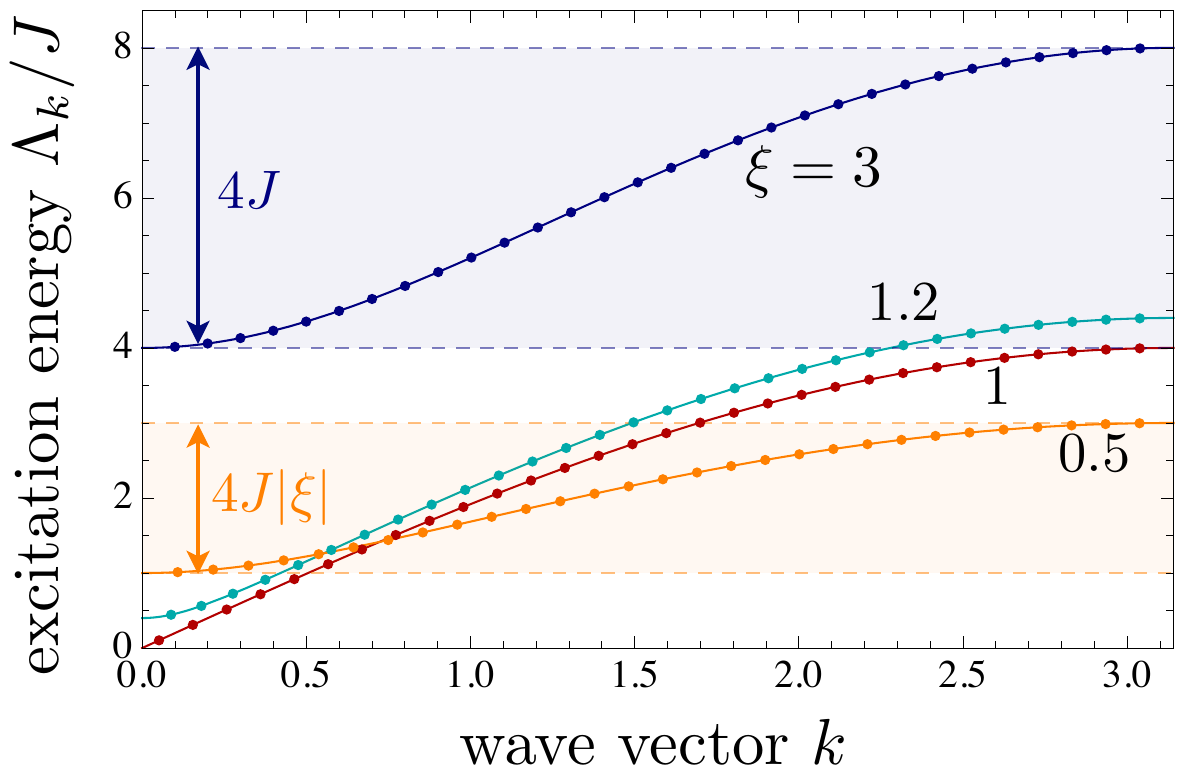} 
	\caption{
	Excitation energies $\Lambda_k$ vs. the allowed real wave vectors $k$ of a transverse-field Ising chain with $N=30$ 	
	and $\xi=0.5,1,1.2,3$ (dots). For $\xi=0.5$, there is also one imaginary wave vector (see text). 
	The solid lines are a guide to the eye. 
	The shaded regions indicate the bandwidth of the Ising chain for $\xi=0.5$ (ferromagnetic phase, orange) 
	and for $\xi=3$ (paramagnetic phase, blue). 
}
	\label{fig:S3}
\end{figure} 
For $N \rightarrow \infty$, $\mathcal{H}_I$ undergoes a second order QPT at 
$\xi =\pm 1$ from a ferromagnetic [$\xi \in (0, 1)$] or an
antiferromagnetic [$\xi \in ( - 1,0)$] ordered phase with doubly degenerate 
eigenstates ($\Lambda_{k^\prime} \rightarrow 0 $) to a paramagnetic disordered phase
($|\xi| >1$) with $\Lambda_k > 0$ for all $k$. This QPT is signaled by correlators of the order
parameter $\sigma_x$. 
Note, though, that $\langle \sigma_x^j \rangle \equiv 0$ for
all $\xi$. Since $\mathcal{H}_I$ commutes with $\prod _j \sigma_z^j$, all eigenstates
of $\mathcal{H}_I$ formally obey this symmetry that maps $\sigma_x^j \rightarrow -
\sigma_x^j$.

Fig.\ \ref{fig:S3} shows the excitation energies $\Lambda_k$ of $\mathcal{H}_I$ vs.
the allowed (real) wave vectors $k$ for $N=30$ and various $\xi$ (for $\xi=0.5$,
there is one imaginary wave vector $k^\prime \approx 0.693 i$, and
$\Lambda_{k^\prime}\approx 0$). In the limit $N\rightarrow \infty$, the $\Lambda_k$
form a continuous band. Its gap is given by $|1-|\xi||$ and vanishes at the quantum
critical point $|\xi|=1$. In the disordered phase ($|\xi|>1$), the bandwidth is $4J$
(indicated for $\xi=3$ in Fig.\ \ref{fig:S3}) and independent of $\xi$. In the
ordered phase ($|\xi|<1$), the bandwidth is given by $4J|\xi|$. 

\begin{figure}
	\centering
	\includegraphics[width=0.8\columnwidth]{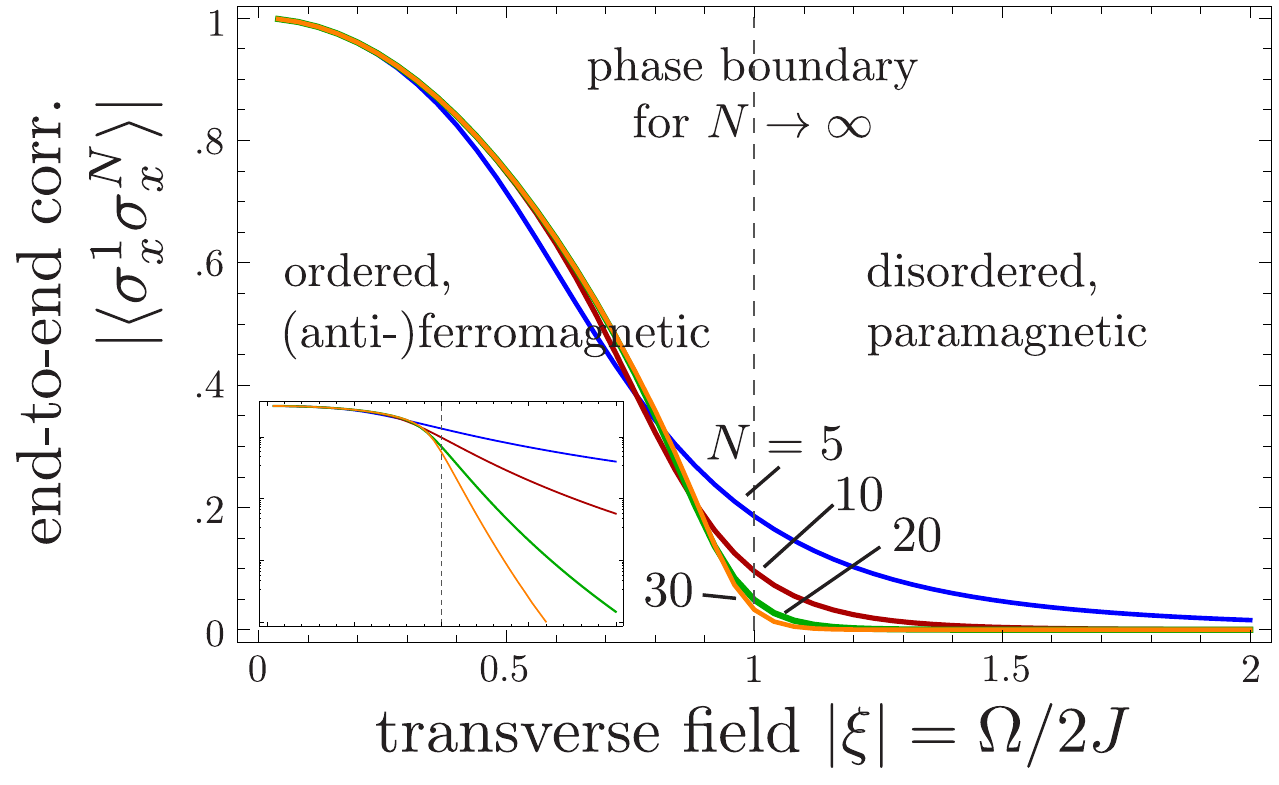} 
	\caption{
	End-to-end correlator $|\langle \sigma_x^1 \sigma_x^N \rangle|$ vs.\ normalized magnetic field $|\xi|=\Omega/2 J$ for $N=5,10,20,30$ (blue, red, green, orange). The signs of $\langle \sigma_x^1 \sigma_x^N \rangle (\xi)$ and $\xi$ agree except that $N$ is odd and $\xi <0$.
	Inset, same plot but $|\langle \sigma_x^1 \sigma_x^N \rangle|$ on a logarithmic scale covering values from $10^{-7}$ to $1$. 
}
	\label{fig:S3a}
\end{figure} 
Signatures of the QPT are already present for relatively small system sizes. This is evident from Fig.\ \ref{fig:S3a} where we plot the end-to-end correlator $\langle \sigma_x^1 \sigma_x^N \rangle$, an order parameter of the QPT for $N\rightarrow \infty$, as function of $\xi$ for different (finite) $N$  (at zero temperature; see Sec.\ V and \cite{Supp_Lieb1961} for calculations). Already for $N \gtrsim 10$, the end-to-end correlator becomes very small at $|\xi| \approx 1$ and displays a distinct transition from algebraic to exponential decay (see inset of Fig.\ \ref{fig:S3}). This illustrates that even small Ising chains of a comparable size exhibit interesting quantum many-body physics. For more details on
the transverse-field Ising chain and its QPT, see, e.g., \cite{Supp_Lieb1961,Supp_Pfeuty1970,Supp_Sachdev1999}.

Assuming zero temperature, the qubit autocorrelator $\rho(t)=\langle \sigma_x^1(t) \sigma_x^1(0) \rangle$ can now be easily calculated using 
\ma{
\sigma_x^1 = c_1^\dagger + c_1 = \sum_k \phi_{k,1} (\eta_k^\dagger + \eta_k).
}
One obtains
\ma{
\rho(t) = \sum_{k} \phi_{k,1}^2 e^{-i t \Lambda_k}.\label{eq:Supplement_autocorrelator}
}
The Fourier transform $\tilde{\rho} (\omega)$ of $\rho(t)$ is a sum of delta peaks. In order to obtain a continuous spectrum $\tilde{\rho}(\omega)$, we have to take the limit $N\rightarrow \infty$ in Eq.\ \eqref{eq:Supplement_autocorrelator}. As its RHS contains rapidly oscillating terms for $N\rightarrow \infty$ (like $\sin Nk$), it cannot be straightforwardly transformed into an integral via a Riemann sum. We therefore write $k_l = \pi/N(l-\nu_l)$ for $l=1,\ldots,N$ \cite{Supp_Lieb1961} and find, by means of Eq.\ \eqref{eq:Supplement_kvalues},
\ma{
\nu_l = \dfrac{1}{\pi} \arctan \Big[ \dfrac{\xi \sin (\pi l/N)}{\xi \cos( \pi l/N) -1} \Big]+ \mathcal{O}(1/N).
}
With these expressions for $k_l$ and $\nu_l$, $\rho(t)$ can be transformed into an integral $\int_1^N \mathrm{d}l$ for $N\rightarrow \infty$. Substituting $\mathrm{d}l\rightarrow \mathrm{d}k$ ($k$ as defined above, $\mathrm{d}k/\mathrm{d}l\approx \pi/N$) and dropping all terms $\mathcal{O}(1/N)$ finally leads to 
\begin{align}
\rho(t) &= \Theta (1-|\xi|) (1-|\xi|^2) \notag\\
&+ \dfrac{2 }{\pi} \int_0^\pi \mathrm{d} k \dfrac{ \xi^2 \sin^2 k}{1+\xi^2 -2\xi \cos k} e^{-it \Lambda(k)} ,\label{eq:Supplement_autocorrcont}
\end{align}
where $\Theta(x)$ is the Heaviside step function and $\Lambda(k)$ stands for $\Lambda_k$ with continuous $k$. The first term on the RHS of Eq.\  \eqref{eq:Supplement_autocorrcont} is the $k^\prime$-term in Eq.\ \eqref{eq:Supplement_autocorrelator} for $N\rightarrow \infty$, which must be treated separately. It causes a nonzero mean value of $\mathrm{Re}\rho(t)$ in the ordered phase. Taking the Fourier transform of Eq.\ \eqref{eq:Supplement_autocorrcont} yields
\ma{
\tilde{\rho}(\omega)&= 2\pi \delta(\omega)\Theta (1-|\xi|) (1-|\xi|^2) \notag\\
&+ \Theta(\omega -2J|1-|\xi||) \;\Theta(2J|1+|\xi||-\omega) \notag \\
& \times \dfrac{4 |\xi|}{\omega} \sqrt{1-\cos^2 k(\omega)}, \label{eq:Supplement_rhoxxomegacont}
}
where $\cos k(\omega) = [1+\xi^2 -(\frac{\omega}{2J})^2]/(2\xi)$. Note that this
result does not depend on the sign of $\mathcal{J}$ (and the sign of $\xi=\Omega/2\mathcal{J}$).
For ferromagnetic coupling $\mathcal{J}>0$ (and $\xi>0$), Eq.\
\eqref{eq:Supplement_rhoxxomegacont} can be simplified to the form of Eq.\ (3) and is
plotted in Fig.\ 2(a) of the main text. For antiferromagnetic coupling
$\mathcal{J}<0$ (and $\xi<0$), one just has to replace $\xi \rightarrow |\xi|$ in
Eq.\ (3). Thus, with this replacement, our discussion of $\tilde{\rho}(\omega)$ below
Eq.\ (3) and the plots in Fig.\ 2 of the main text hold for antiferromagnetic coupling as well.

\section{III. Spectrum of the resonator}

In this section, we calculate the spectrum $S (\omega)$ of the resonator of our
system, which is coupled to the Ising chain. Complementary to Figs.\ 2(b) and
2(c) of the main text, we plot $S(\omega)$ in the limiting cases $g/J \ll 1$ and $g/J
\gg 1$, and for finite $N$. In these plots, we vary the transverse field $\xi$ at
fixed qubit-qubit coupling $J$, as experimentally realistic for Cooper-pair boxes 
(see Sec.\ I  of these supplementary notes).
However, if the proposed setup is implemented with standard transmons instead of Cooper-pair
boxes, then $\xi$ will be constant and $J$ will be flux-tunable. We 
also provide plots of $S(\omega)$ for this scenario. We remark that, like $\tilde{\rho}(\omega)$,
$S(\omega)$ turns out to be independent of the sign of $\mathcal{J}$ (and of the sign
of $\xi = \Omega/2\mathcal{J}$). For ease of notation, we will therefore refer to $J$ as the qubit-qubit
coupling and identify $\xi = |\xi|=\Omega/2J$ where appropriate throughout this section.

In order to calculate $S(\omega)$, we assume $g/\omega_0 \ll 1$ and linearize the
Hamiltonian $\mathcal{H}$ [Eq.\ (1) of the main text]. That is, we now consider the
Hamiltonian 
\ma{
\tilde{\mathcal{H}} = \dfrac{1}{2}(p_0^2 + \omega_0^2 x_0^2) + x_0 \sum_{j=1}^N \lambda_j x_j + \mathcal{H}_h, \label{eq:Supplement_simplifiedmodel}
}
where $\lambda_j= \sqrt{2 g^2 \omega_0} \tilde{\lambda}_j$ and $\mathcal{H}_h=\sum_{j=1}^N(p_j^2+w_j^2x_j^2)/2$. It is obtained by substituting the coupling term in Eq.\ (1) by $g(a^\dagger + a) \sum_{j=1}^N \tilde{\lambda}_j x_j$ ($\tilde{\lambda}_j$ is a coupling constant) and $\mathcal{H}_I$ by $\mathcal{H}_h$, the Hamiltonian of a set of harmonic oscillators with frequencies $w_j$, and by introducing canonical coordinates for the resonator via $ x_0 = 1/\sqrt{2 \omega_0}(a^\dagger +a)$ and $p_0=i \sqrt{ \omega_0/2}(a^\dagger - a)$. Note that $x_0$ couples to a force $F_h(t) = \sum_{j=1}^N \lambda_j x_j(t)$ in Eq.\ \eqref{eq:Supplement_simplifiedmodel}. By writing $\mathcal{H}$ [Eq.\ (1)] in terms of $x_0$ and $p_0$, one finds that here $x_0$ couples to a force $F_I(t) = \sqrt{2 g^2 \omega_0} \sigma_x^1(t)$. The parameters $\tilde{\lambda}_j$ and $w_j$ in $\tilde{\mathcal{H}}$ can be chosen such that 
\ma{ 
\langle F_h(t) F_h(0) \rangle = \sum_{j=1}^N \dfrac{ \lambda_j^2}{2 w_j} e^{- i w_j t}
=  \langle F_I(t) F_I(0) \rangle \label{eq:Supplement_spectrum_bath}
}
(in this case also the spectra of the forces will agree). Indeed, $w_j=\Lambda_{k_j}$ and $\tilde{\lambda}_j^2 = 2 w_j A_{k_j}^2 \sin^2 N k_j$ guarantee Eq.\ \eqref{eq:Supplement_spectrum_bath}. We now calculate $S (\omega)$, the Fourier transform of $2 \omega_0 \langle \tilde{0}| x_0(t) x_0 |\tilde{0} \rangle$, where $\ket{\tilde{0}}$ is the ground state of $\tilde{\mathcal{H}}$. To that end, we first reformulate
\ma{
\tilde{\mathcal{H}} = \dfrac{1}{2} (\mathbf{P}^T \mathbf{P} + \mathbf{X}^T \underline{\Omega}^2 \mathbf{X}),
}
with $\mathbf{X}^T = (x_0,x_1,\ldots, x_N)$, $\mathbf{P}^T = (p_0,p_1,\ldots, p_N)$, and
\begin{align}
\underline{\Omega}^2 = 
\left(
   \begin{array}{cccc}
     \omega_0^2 & \lambda_1 & \ldots & \lambda_N\\
     \lambda_1 & w_1^2 & & \\
     \vdots &  & \ddots &  \\
     \lambda_N & &  & w_N^2 \\ 
   \end{array}
\right). \label{eq:Supplement_Omega^2}
\end{align}
There is an orthogonal matrix $G$ for which 
\ma{
\tilde{\mathcal{H}} = \dfrac{1}{2} (\tilde{\mathbf{P}}^T \tilde{\mathbf{P}} + \tilde{\mathbf{X}}^T \underline{\tilde{\Omega}}^2 \tilde{\mathbf{X}}),
}
where $\tilde{\mathbf{X}} = G^T \mathbf{X} $, $\tilde{\mathbf{P}} = G^T \mathbf{P} $, and $\underline{\tilde{\Omega}}^2$ is diagonal with 
$\tilde{\Omega}_j^2 \equiv (\underline{\tilde{\Omega}}^2)_{jj}$ being an eigenvalue of $\underline{\Omega}^2$. We calculate 
\ma{
\langle \tilde{0}| x_0(t) x_0 |\tilde{0} \rangle & = \sum_{j,j^\prime =0 }^{N} G_{0,j} G_{0, j^\prime} 
\langle \tilde{0}| \tilde{x}_j(t) \tilde{x}_{j^\prime} |\tilde{0} \rangle \\
&= \sum_{j=0}^ {N} \dfrac{G_{0,j}^2}{2 \tilde{\Omega}_j} e^{-i t \tilde{\Omega}_j}
}
and obtain with that
\ma{
S (\omega)& = 2 \pi \omega_0 \sum_{j=0}^ {N}   \dfrac{G_{0,j}^2}{\tilde{\Omega}_j} \delta(\omega - \tilde{\Omega}_j)\\
&= 4 \Theta(\omega) \omega_0 \mathrm{Im} \big[ \mathcal{R} (\underline{\Omega}^2, \omega^2 - i 0^+)_{0,0} \big].
}
In the last line, the matrix element of the resolvent $\mathcal{R}( \underline{\Omega}^2, \omega^2 ) = (\omega^2 - \underline{\Omega}^2)^{-1}$
is to be taken in the basis in that $\underline{\Omega}^2$ has the form of Eq.\ \eqref{eq:Supplement_Omega^2}. It can be calculated following Ref.\ \cite{Supp_Marquardt2005}. The result is
\begin{gather}
S  (\omega) =  
\dfrac{ 4 \Theta (\omega) \omega_0  [\langle F_h F_h \rangle_\omega/2 + 0^+]}
{[\omega^2 - \omega_0^2 - 2\tilde{\chi}(\omega^2 ) ]^2\!+\![0^+\! + \langle F_h F_h \rangle_\omega/2  ]^2}, \\
\tilde{\chi}(\omega^2 ) = \dfrac{1}{2\pi} \int d\Omega \dfrac{ \Omega  \langle F_h F_h \rangle_\Omega}{\omega^2 - \Omega^2}.
\end{gather}
Thus, we have expressed the spectrum of the resonator $S (\omega)$ in terms of the spectrum $\langle F_h F_h \rangle_\omega $ of the bath of harmonic oscillators which is the Fourier transform of $\langle F_h(t) F_h(0) \rangle$ [Eq.\ \eqref{eq:Supplement_spectrum_bath}]. Note that in the limit $N\rightarrow \infty $, $\langle F_h F_h \rangle_\omega $ can become continuous and then $\tilde{\chi}(\omega^2 )$ is a principal value integral. If we now assume that we have chosen $\tilde{\lambda}_j$ and $w_j$ in $\tilde{\mathcal{H}}$ [Eq.\ \eqref{eq:Supplement_simplifiedmodel}] such that Eq.\ \eqref{eq:Supplement_spectrum_bath} holds, we can substitute $\langle F_h F_h \rangle_\omega \rightarrow \langle F_I F_I \rangle_\omega = 2g^2 \omega_0 \tilde{\rho} (\omega)$. This leads to
\ma{
S (\omega) \!=\!  \dfrac{4 \Theta (\omega) \omega_0 [g^2 \omega_0 \tilde{\rho}(\omega) + 0^+ ]}{[\omega^2\!-\!\omega_0^2\!-\!4g^2 \omega_0 \chi(\omega^2)]^2\!+\![0^+ \! + g^2 \omega_0  \tilde{\rho}(\omega) ]^2}, \label{eq:Supplement_S_coupled_0}
}
where $\chi(\omega^2) $ is the principal value integral
\ma{
\chi(\omega^2) = \dfrac{1}{2\pi} \int \mathrm{d} \Omega \dfrac{\tilde{\rho}(\Omega) \Omega}{\omega^2-\Omega^2}.
} 
Note that $S (\omega) |_{g=0}= 2 \pi \delta ( \omega-\omega_0)$. However, the
spectrum of any realistic microwave resonator at $g=0$ will be a Lorentzian with full
linewidth $ \kappa$ at half maximum. We use the case $g=0$ to relate the so far
infinitesimal real number $0^+$ in Eq.\ \eqref{eq:Supplement_S_coupled_0} to $\kappa$
by demanding
\ma{
S (\omega) |_{g=0} =  \dfrac{4 \Theta (\omega) \omega_0 0^+ }{(\omega^2 - \omega_0^2 )^2 + (0^+)^2 } \approx
\dfrac{\kappa}{  ( \omega-\omega_0)^2 + (\kappa/2)^2}.  \label{eq:Supplement_finite_kappa}
}
For $\kappa \ll \omega_0$, it is sufficient to focus on the vicinity of the strongly
pronounced peak of $S (\omega) |_{g=0}$ at $\omega=\omega_0$ (i.e., on
$\omega-\omega_0 \ll \omega_0$), and we find that here Eq.\
\eqref{eq:Supplement_finite_kappa} is fulfilled for $0^+ =  \kappa \omega_0 $.
Inserting this expression in Eq.\ \eqref{eq:Supplement_S_coupled_0} finally leads to
Eq.\ (4) of the main text. Note that the properties of the TFIC enter 
our result for $S(\omega)$ only via the spectrum $\tilde{\rho}(\omega)$ of the bare
TFIC. Therefore, Eq.\ (4) also holds if the resonator is coupled to a
different system than the TFIC, with some other spectrum. Note further that
$S(\omega)$ is independent of the sign of $\mathcal{J}$ (and the sign of
$\xi=\Omega/2\mathcal{J}$) because the spectrum $\tilde{\rho}(\omega)$ of the TFIC has
this property.

\begin{figure}
	\centering
	\includegraphics[width=1\columnwidth]{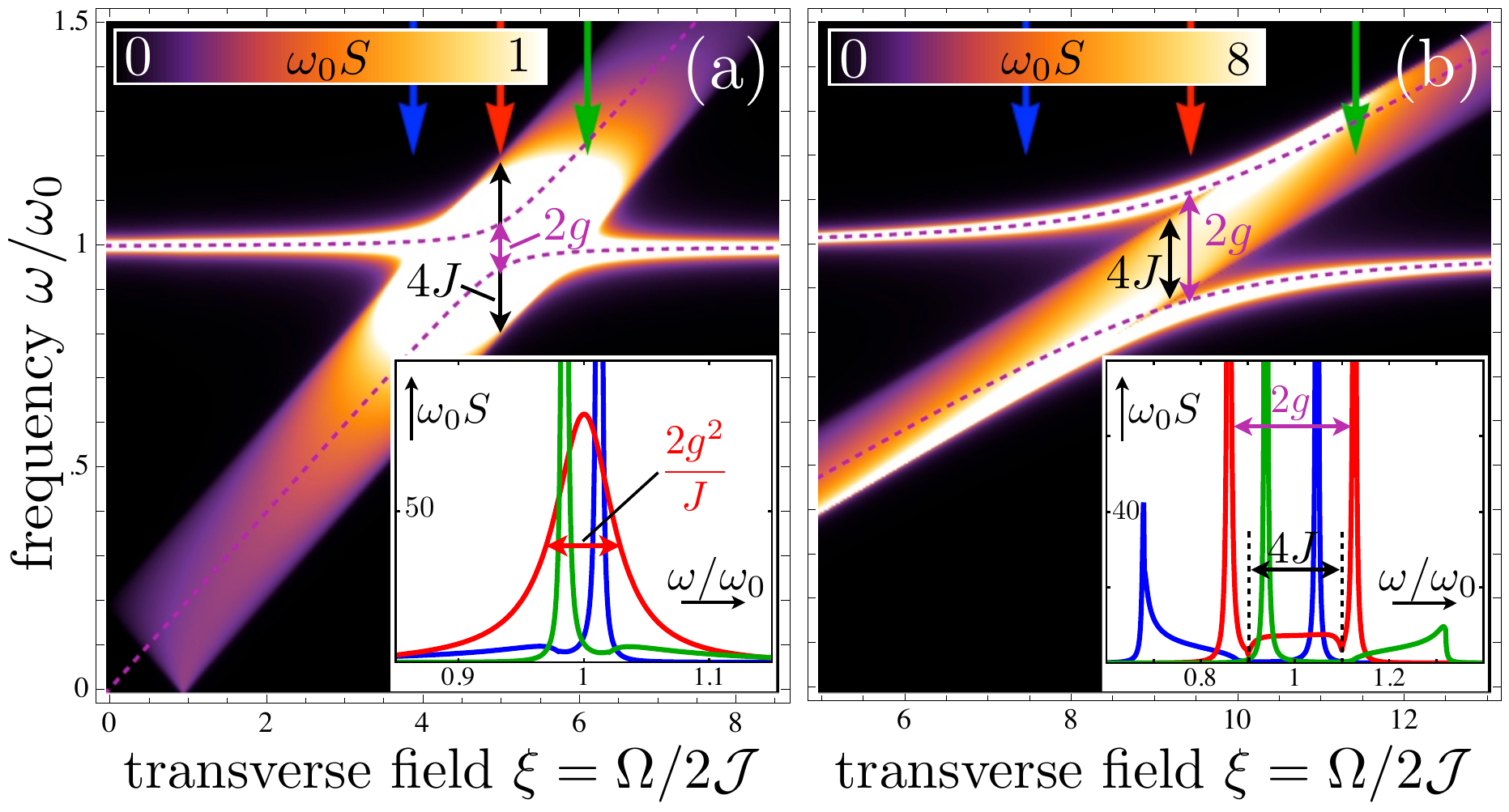}
	\caption{
	Spectrum $ S $ of a resonator coupled to the first spin of an Ising chain 
	($N \rightarrow \infty$) vs.\ frequency $\omega$ and normalized transverse field $\xi$.
	(a) The case $g/J\ll 1$ (the parameters are $g=0.05$, $J=0.1$, and $\kappa = 10^{-4}$). 
	Inset, $S(\omega)$ for $\xi =3.9,5,6.1$ (blue, red, green). 
	(b) The case $g/J\gg 1$ (the parameters are $g=0.12$, $J=0.05$, and $\kappa = 10^{-4}$). 
	Inset, $S(\omega)$ for $\xi =7.8,10,12.2$ (blue, red, green). 
	All parameters are measured in units of $\omega_0$. 
	The dashed lines are the first two excitation energies of $\mathcal{H}$ for the same parameters, but $N=1$. 
	For better visibility of the features, values of $S(\omega)> 1$ [$S(\omega)>8$] in the density plot of 
	(a) [(b)] are plotted in white.
	The lines in the insets correspond to cuts along the arrows in the main plots. }
	\label{fig:S4}
\end{figure} 
Fig.\ \ref{fig:S4} complements Figs.\ 2(b) and 2(c) of the main text by showing $S
(\omega) $ in the limiting cases $g/J\ll 1$ [Fig.\ \ref{fig:S4}(a)] and $g/J\gg 1$
[Fig.\ \ref{fig:S4}(b)]. In Fig.\ \ref{fig:S4}(a), we choose the parameters
$J/\omega_0$ and $\kappa/\omega_0$ as in Fig.\ 2(b), but $g/\omega_0 = 0.05$. Where
the Ising chain is off-resonant with $\omega_0$, the spectrum is qualitatively
similar to the one of Fig.\ 2(b). Also here one observes the dispersive shift
($\propto g^2$) of the resonator frequency in analogy to the $N=1$ case and a broad
side maximum of width $\sim 4J$ (blue and green lines in the inset). Both are less
pronounced than in Fig.\ 2(b) due to the lower value of $g$. On resonance ($\xi
\approx \omega_0/2J$), though, the double peak structure reminiscent of the $N=1$
case is no longer visible. Instead, $S(\omega)$ is a Lorentzian around $\omega_0$
with full width at half maximum given by $2g^2/J$ (as long as $\kappa\ll g^2/J$ and
$\omega$ is within the band of the Ising chain). Indeed, assuming small $g/J$, one
may replace $\tilde{\rho}(\omega)$ by its maximum $2/J$ and take $\chi(\omega^2)
\approx 0$ in Eq.\ \eqref{eq:Supplement_S_coupled_0}. One can then verify
\ma{
S(\omega)|_{ \xi\approx \omega_0/2J} \approx \dfrac{2 g^2/J}{(\omega-\omega_0)^2 +(g^2/J)^2}.
}
In Fig.\ \ref{fig:S4}(b), we choose the parameters $g/\omega_0$ and $\kappa/\omega_0$
as in Fig.\ 2(b), but $J/\omega_0 = 0.05$. This case has already much similarity with
the usual single-qubit case. Off resonance, the resonator experiences again the same
dispersive shift as for $N=1$. On resonance, the broad double peak structure of
Figs.\ 2(b,c) with width $4J$ has developed into two sharp Lorentzians separated by
$\approx 2g$ as for $N=1$ (red line in the inset). The chain is visible only as faint
band of width $4J$ in between these peaks. 
\begin{figure}
	\centering
	\includegraphics[width=1\columnwidth]{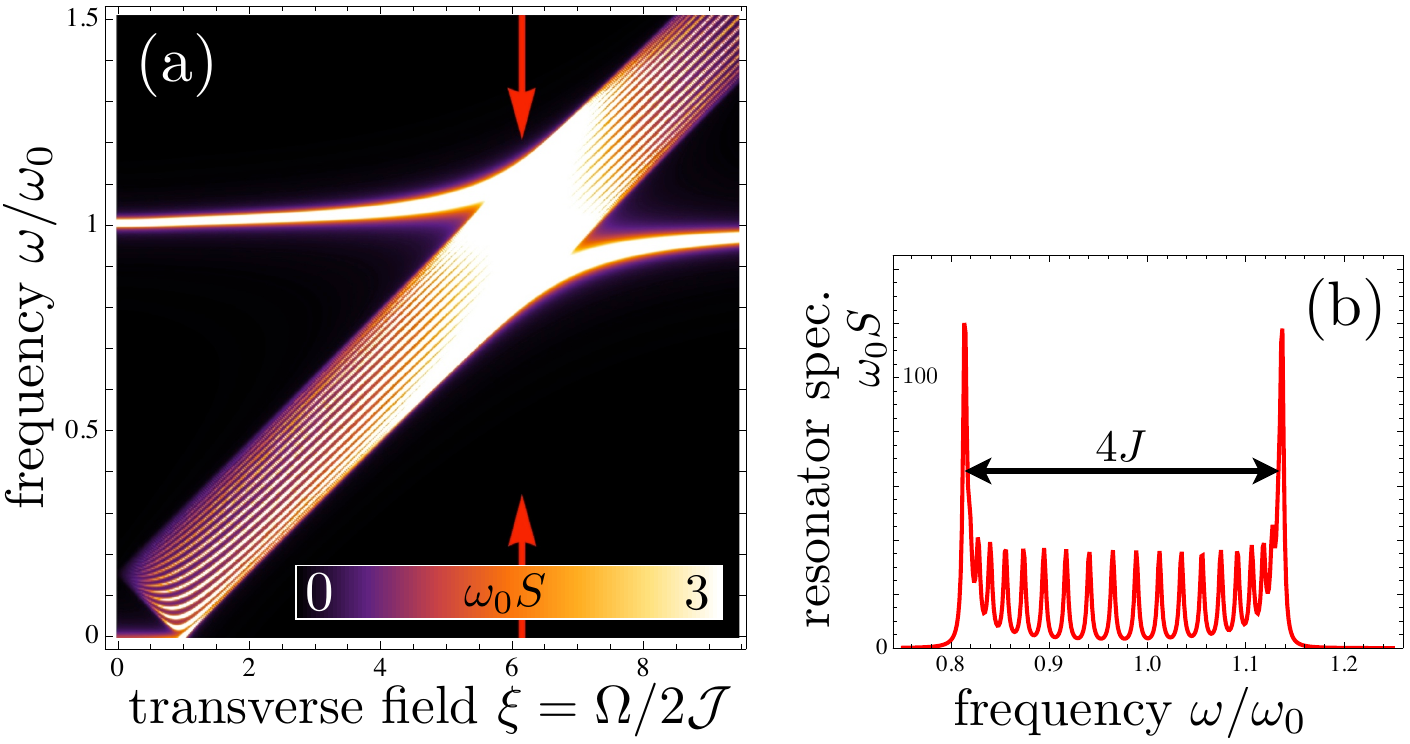}
	\caption{
	(a) Spectrum $ S $ of a resonator coupled to the first spin of a finite Ising chain ($N=20$) 
	vs.\ frequency $\omega$ and normalized transverse field $\xi$.
	The parameters are $g=0.12$, $J=0.08$, $ \kappa = 10^{-4}$, and $ \gamma = 5
    \times 10^{-3}$ (in units of $\omega_0$). 
	(b) $S(\omega)$ for $\xi =6.1$. This curve corresponds to a cut along the arrows in (a).
	}
	\label{fig:S5}
\end{figure} 

In order to illustrate finite-size effects on the resonator spectrum $S(\omega)$, we
calculate the spectrum $\tilde{\rho}(\omega)$ of a finite transverse-field Ising
chain. It is given by the Fourier transform of Eq.\
\eqref{eq:Supplement_autocorrelator} and reads $\tilde{\rho}(\omega) = 2 \pi
\sum_{k} \phi_{k,1}^2 \delta(\omega- \Lambda_k)$. We assume that the delta peaks in
$\tilde{\rho}$ are broadened by decay processes and replace them with Lorentzians
centered around $\Lambda_k$ and having a full width at half maximum of $\gamma$.
Together with Eq.\ (4) of the main text, this yields the spectrum $S(\omega)$ of a
resonator coupled to a TFIC of finite length. In Fig.\ \ref{fig:S5}, we plot
$S(\omega)$ for similar system parameters as in Fig.\ 2 of the main text ($g/J
\approx 1$), but $N=20$. Signatures of the QPT at $\xi=1$, the dispersive shift of
the resonator frequency, and the double-peak structure on resonance with $4J$
separation of the peaks (rather than $2g$ as in the case $N=1$) are present also for
$N=20$. We remark that compared to the case $N\rightarrow \infty$ (Fig.\ 2), the
ratio $g/J$ has to be slightly increased for $N=20$ (Fig.\ \ref{fig:S5}) such that
the double peak structure of $S(\omega)$ on resonance is clearly visible. This is
because the weight of the edges of the band of the Ising chain in the spectrum
$\tilde{\rho}(\omega)$ increases with $N$.

Finally, we plot $S(\omega)$ for varying values of the qubit-qubit coupling
$J$ and keep the normalized transverse field $\xi= \Omega/2J $ constant. This corresponds to
an implementation of our proposal with usual flux-tunable transmons.
In such an implementation, $J$ and $\Omega$ change with the external flux approximately
in the same proportion. Thus, $J$ is tunable and $\xi$ is constant 
(see Sec.\ I of these supplementary notes). 

\begin{figure}
	\centering
	\includegraphics[width=1\columnwidth]{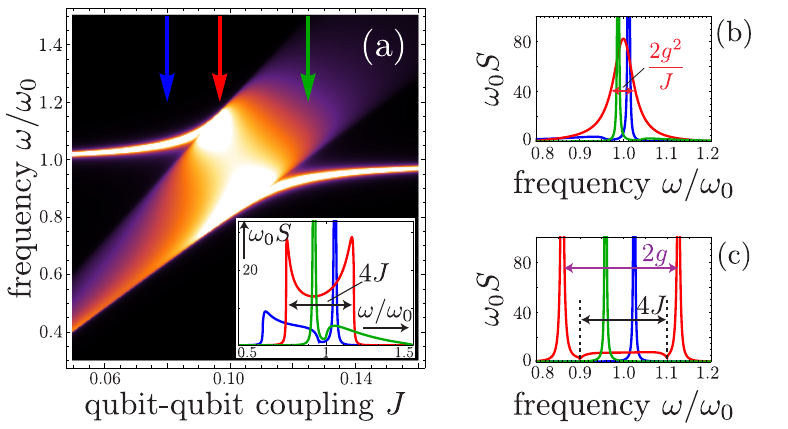}
	\caption{(a) Spectrum $S$ of a resonator coupled to the first spin of an Ising chain
    ($N\rightarrow \infty$) vs. probe frequency $\omega$ and qubit-qubit coupling
    $J$. The normalized transverse field $\xi$ is constant ($\xi=5$). 
    This corresponds to an implementation of the Ising chain with
    standard transmons. Here, the resonator and the first spin couple with a strength $g=0.12$.
    The color scale covers values of $S$ from $0$ (black) 
    to $15$ (white), and values $>15$ are also plotted in white. Inset, $S(\omega)$ for
    the same parameters and $J=0.08,0.096,0.125$ (blue, red, green). These curves
    correspond to cuts along the arrows through the density plot of (a).
    (b) Spectrum $S$ as in (a) in the limiting case $g/J \ll1$. The plot shows
    $S(\omega)$ for $\xi=5$, $g=0.05$, and $J=0.08,0.1,0.13$ (blue, red,
    green). (c) Spectrum $S$ as in (a) in the limiting case $g/J \gg 1$. The plot shows
    $S(\omega)$ for $\xi=10$, $g=0.12$, and $J=0.03,0.05,0.07$ (blue, red, green).
    For all plots we have chosen the resonator linewidth $\kappa = 10^{-4}$.
    All parameters are measured in units of the resonator
    frequency $\omega_0$.
	}
	\label{fig:S6}
\end{figure} 
An Ising chain with tunable $J$ but constant $\xi$ is confined to one phase.
If implemented with transmons, this has to be the
paramagnetic phase ($\xi>1$; see Sec.\ I).
Thus, when plotted as function of $J$
at constant $\xi$, the resonator spectrum $S(\omega)$ will not carry signatures of a phase
transition. Moreover, the bandwidth of the chain 
($4J\xi$ for $\xi<1$ and $4J$ for $\xi>1$) will not be constant. Otherwise $S(\omega)$
displays the same features for transmons as before for CPBs, as Fig.\ \ref{fig:S6}
demonstrates. Before discussing Fig.\ \ref{fig:S6}, we remark that 
the tunability of $J$ for transmons implies that ratio $g/J$ is
not constant. We have seen that shape of the spectrum $S$ depends crucially on the ratio $g/J$ 
if the Ising chain is resonant with the resonator. Therefore, we differentiate the 
cases $g/J \ll 1$, $g/J \approx 1$, and $g/J \gg 1$ (as for CPBs) for the
Ising chain formed by transmons being resonant with the resonator frequency
$\omega_0$. 

Under these
conditions, Fig.\ \ref{fig:S6}(a) corresponds to
Figs.\ 2(b) and 2(c) of the main text. That is, these figures illustrate the situation where the
qubit-qubit coupling $J$ in a semi-infinite chain of transmons resonant with
$\omega_0$ [Fig.\ \ref{fig:S6}(a)] and in a semi-infinite chain of CPBs
[Figs.\ 2(b,c)] is comparable to the coupling $g$ of the respective first artificial atom and the resonator. 
Explicitly, like in Figs.\ 2(b,c), we have chosen $g/\omega_0=0.12$ in Fig.\
\ref{fig:S6}(a). Moreover, the choice $\xi=5$
(a realistic value for transmons) ensures that the center of the band of the Ising chain
($2J\xi$) formed by transmons is on resonance with the resonator at $J/\omega_0 =
0.1$ [like in Figs. 2(b,c)]. As expected, the bandwidth of the TFIC increases linearly with $J$ in Fig.\
\ref{fig:S6}(a). Out of
resonance, one observes the usual dispersive shift of the resonator frequency. On
resonance, the spectrum exhibits the characteristic double-peak structure with $4J$
separation of the peaks, which is also present for Cooper-pair boxes [Fig.\ 2(c)].

Also in the limiting cases $g/J \ll1$ and $g/J \gg 1$ (on resonance), a chain of
transmons displays the same behavior that we have found
before for CPBs: Fig.\ \ref{fig:S6}(b) shows $S(\omega)$ for $\xi=5$ as in
(a), but with $g/\omega_0 = 0.05$. For the different curves, $J$ is chosen such that
the TFIC is below (blue), on resonance with (red), and above (green) the resonator
frequency $\omega_0$. This plot corresponds to the inset of Fig.\ \ref{fig:S4}(a).
The spectrum of a chain of transmons weakly coupled to a resonator is essentially
identical to the one for a chain of CPBs, and its features can be explained in the
same manner. In order to study the limiting case $g/J \gg 1$ for transmons,
we chose $g=0.12$ and $\xi=10$ for the curves in Fig.\ \ref{fig:S6}(c). With this choice of $\xi$, the
Ising chain formed by transmons is on resonance with the resonator at $J=0.05$. This
was also the case in Fig.\ \ref{fig:S4}(b), where we have studied the
limiting case $g/J \gg 1$ for CPBs. As Fig.\ \ref{fig:S6}(b), Fig.\ \ref{fig:S6}(c) shows
$S(\omega)$ for $J$ chosen such that the TFIC is below (blue), on resonance with (red),
and above (green) the resonator frequency $\omega_0$. Like for CPBs, one can clearly
see how the usual Jaynes-Cummings spectrum (corresponding to the case $N=1$) emerges as
limiting case.

\section{IV. Propagation of a localized excitation in the Ising chain}

\begin{figure}
	\centering
	\includegraphics[width=1\columnwidth]{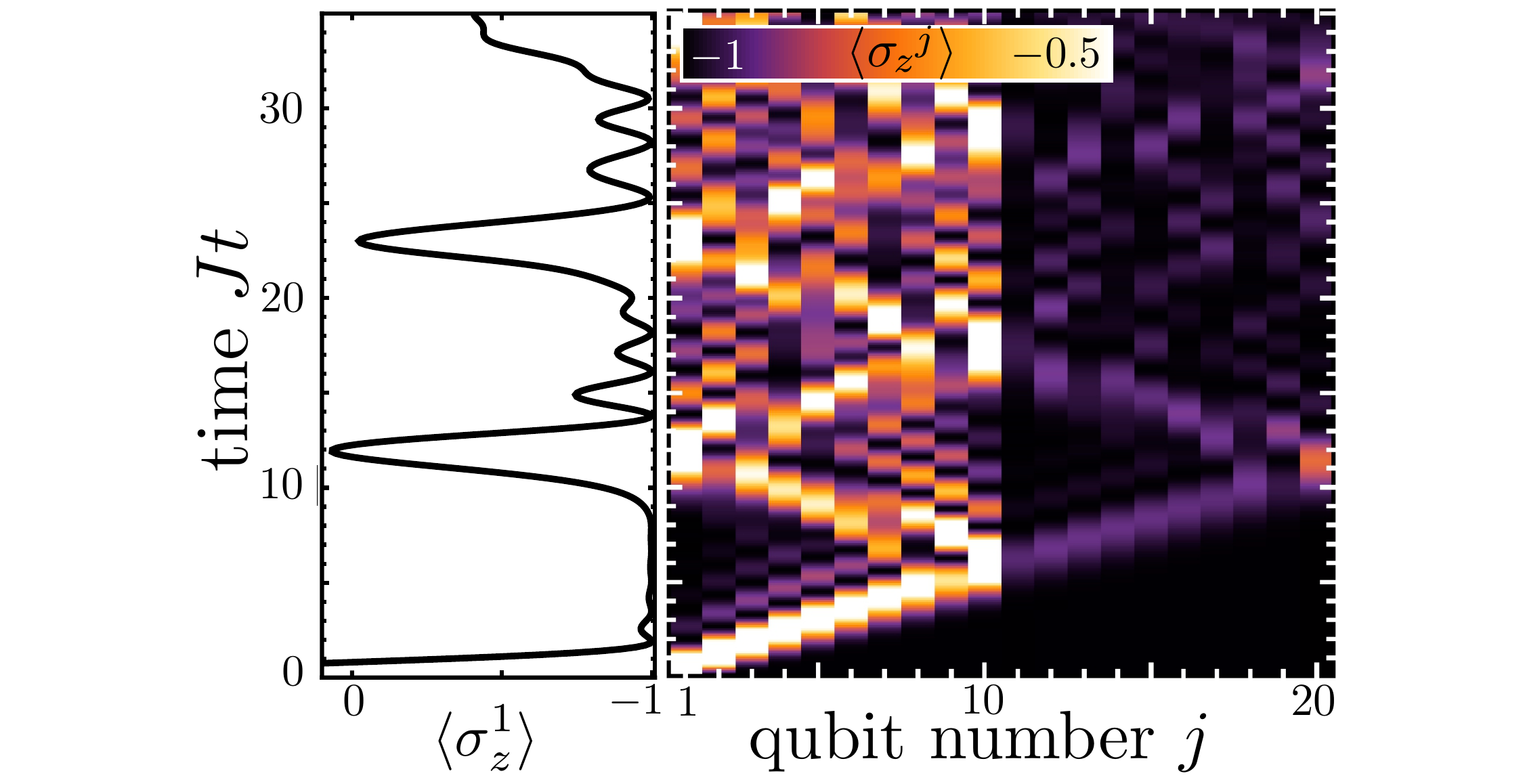}
	\caption{Nonequilibrium time evolution of $\langle \sigma_z^j \rangle $ after a $\pi$-pulse on the first qubit in a 
	transverse-field Ising chain of length $N=20$ in the paramagnetic phase (normalized transverse field $\xi = 8$). 
	Qubit $11$ is strongly detuned from the rest of the chain. Values $>-0.5$ are plotted in white. The measurable observable $\langle \sigma_z^1 \rangle $ is
	singled out left.
	}
	\label{fig:S7}
\end{figure}
This section contains the explicit evaluation of the RHS of Eq.\ (5) of the main
text. Further, it is shown that by deliberately detuning the transition frequency of
one qubit, the effective length of the TFIC can be modified. With 
\ma{
L_j & \equiv c_j^\dagger + c_j = \sum_{k} \phi_{k,j} (\eta_k^\dagger + \eta_k)\\
M_j & \equiv c_j^\dagger -c_j = \sum_{k} \psi_{k,j} (\eta_k^\dagger -  \eta_k),
}
where $\phi_{k,j}$ and $\psi_{k,j}$ are determined by Eqs.\
\eqref{eq:Supplement_relpsiphi} [and explicitly given in Eqs.\
\eqref{eq:Supplement_phi_explicit} and \eqref{eq:Supplement_psi_explicit}], we
reformulate Eq.\ (5) in terms of fermions,
\ma{
\langle \sigma_z^j \rangle (t) = \bra{0} L_1 M_j(t) L_j(t) L_1 \ket{0}. 
}
The RHS of this equation can be evaluated using Wick's theorem, which was first used in this context in Ref.\ \cite{Supp_Lieb1961}. One finds
\ma{
\langle \sigma_z^j \rangle (t) =
& - \sum_k \psi_{k,j} \phi_{k,j} + \sum_{k, k^\prime} e^{i (\Lambda_k - \Lambda_{k^\prime})t} \big[  \phi_{k,1} \phi_{k^\prime,1} \notag \\
&\times  (  \psi_{k,j} \phi_{k^\prime,j} + \psi_{k^\prime,j} \phi_{k,j}) \big]. \label{eq:Supplement_spin_wave_explicit}
}
This formula was used for the plots in Fig.\ 3 of the main text. 

If the transition frequencies $\Omega_j$ of the qubits can be tuned individually, one
can intentionally detune one qubit from the rest of the chain and observe how the
system dynamics changes depending on the detuning. Fig.\ \ref{fig:S7} shows the time
evolution of $\langle \sigma_z^j \rangle (t)$ after a local excitation has been
created on the first site for the same system parameters as in Fig.\ 3, but with
qubit $11$ strongly detuned from the others, explicitly $\Omega_{11} = 1.3 \Omega_j$
for $j \neq11$. This local inhomogeneity acts as a barrier for the propagating
excitation and leads to its reflection. The revival of the measurable observable
$\langle \sigma_z^1 \rangle (t)$ takes place at $t \approx N/2J$ rather than at $t
\approx N/J$ as in Fig.\ 3 of the main text. Thus, strongly detuning one qubit from
the others effectively changes the length of the chain.

\section{V. Quench dynamics of the magnetization and the end-to-end correlations}\label{Sec:Quenches}
We calculate the time evolution of $\langle \sigma_z^j \rangle$ and $\langle
\sigma_x^1 \sigma_x^N \rangle$ that follows a sudden change from $\xi = \xi_a$ to
$\xi=\xi_b$ at $t=0$. We plot and discuss the result for $\langle \sigma_x^1
\sigma_x^N \rangle$ and provide a plot of $\langle \sigma_z^j \rangle$ in addition to
Fig.\ 4 of the main text. In the following, quantities belonging to
$\mathcal{H}_{I,a}$ are labelled by $a$ (like $\Lambda_k^a$), and analogously for
$\mathcal{H}_{I,b}$. 

\begin{figure}
	\centering
	\includegraphics[width=0.7\columnwidth]{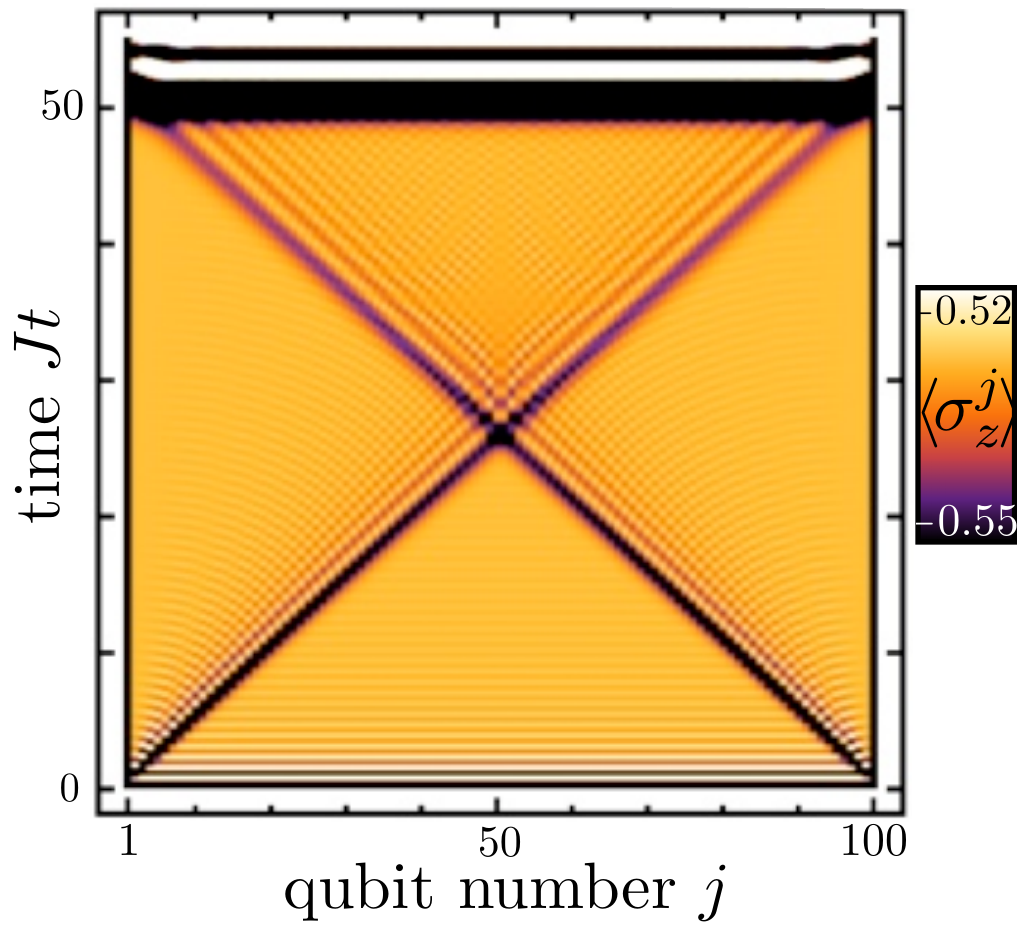}
	\caption{Time evolution of $\langle \sigma_z^j \rangle$ in a transverse-field Ising chain of length $N=100$ 
	after a quench of the normalized transverse field $\xi = 8 \rightarrow 1$. Values $<-0.55$ ($>-0.52$) are plotted in black (white).}
	\label{fig:S8}
\end{figure}

First, we focus on 
\ma{
\langle \sigma_z^j \rangle (t) = \; _a\!\bra{0} e^{i \mathcal{H}_{I,b} t} \sigma_z^j e^{-i \mathcal{H}_{I,b} t} \ket{0}_a.
}
To evaluate the RHS, we use the usual mapping to free fermions
\cite{Supp_Lieb1961,Supp_Pfeuty1970}: We express $\sigma_z^j$ by $\eta_k^b$ and
$\eta_k^{b\dagger}$ whose time dependence is trivial. Then we express these operators
by $\eta_k^a$ and $\eta_k^{a\dagger}$ whose action on $\ket{0}_a$ is known. One
obtains
\begin{align}
\langle &\sigma_z^j \rangle (t) = -\sum_k \psi_{k,j}^b \phi_{k,j}^b + 2 \sum_{k,k^\prime} \lbrace \psi_{k,j}^b \phi_{k^\prime,j}^b \times \notag \\
&[
X_{k,k\prime} \cos t(\Lambda_k^b + \Lambda_{k^\prime}^b) + 
Y_{k,k\prime} \cos t(\Lambda_k^b - \Lambda_{k^\prime}^b)
] \big\rbrace .
\end{align}
Here, 
\ma{
X_{k,k\prime}&=\big[(g_k^b)^T H^a + (h_k^b)^T G^a \big] \big[ (G^a)^T g_{k^\prime}^b + (H^a)^T h_{k^\prime}^b\big], \notag \\
Y_{k,k\prime}&=\big[ (g_k^b)^T H^a + (h_k^b)^T G^a \big] \big[ (H^a)^T g_{k^\prime}^b + (G^a)^T h_{k^\prime}^b \big], \notag
}
and $G$ and $H$ are matrices containing the $g_k$ and $h_k$ as columns, respectively.
Complementary to Fig.\ 4 of the main text, we plot in Fig.\ \ref{fig:S8} $\langle
\sigma_z^j \rangle (t)$ for a quench $\xi_a =8 \rightarrow \xi_b=1$ in a
transverse-field Ising chain with length $N=100$. We focus here on $t \leq T$ and
choose a relatively large chain to strongly contrast the initial approach of $\langle
\sigma_z^j \rangle$ to a constant value with the effects of the finite system size.
The choice of the non-generic value $\xi_b=1$ minimizes dispersion of $\Lambda_k$
and, thus, of the velocities of the quasiparticles. The features of $\langle
\sigma_z^j \rangle (t)$ for $t \leq T$ described in the main text are more pronounced
and clearly visible in Fig.\ \ref{fig:S8}. 

\begin{figure}
	\centering
	\includegraphics[width=\columnwidth]{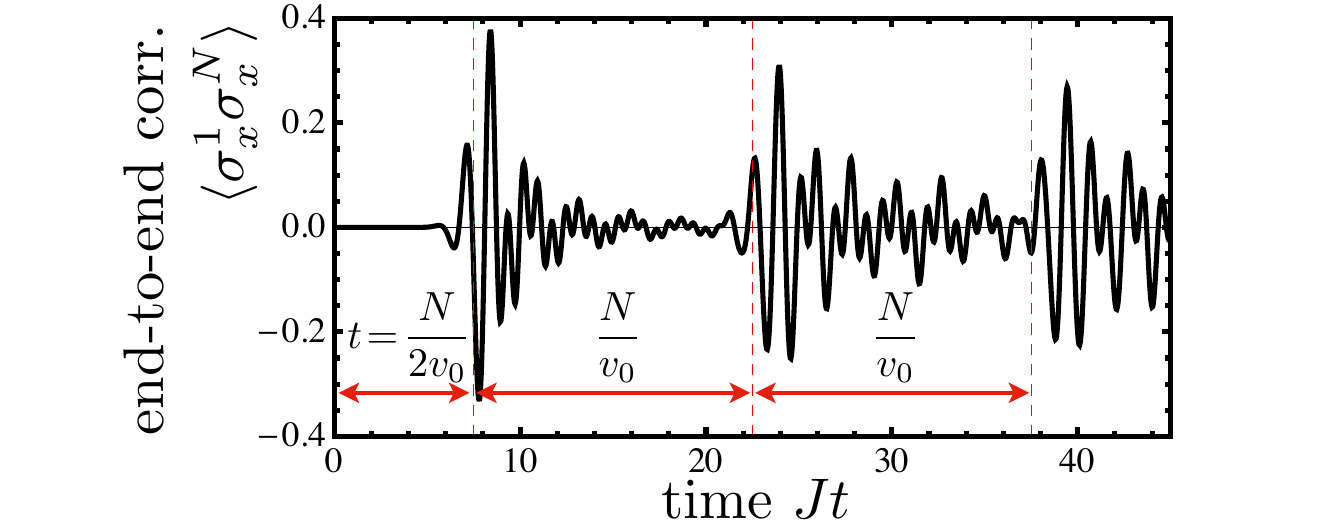}
	\caption{Time evolution of the end-to-end correlator $\langle \sigma_x^1 \sigma_x^N \rangle$ in a transverse-field Ising chain 
	of length $N=30$ after a quench of the normalized transverse field $\xi = 8 \rightarrow 1.5$.
	}
	\label{fig:S9}
\end{figure}

Let us now turn to the quench dynamics of the end-to-end correlator
\ma{
\langle \sigma_x^1 \sigma_x^N \rangle (t) =  \; _a\!\bra{0} e^{i \mathcal{H}_{I,b} t} \sigma_x^1 \sigma_x^N e^{-i \mathcal{H}_{I,b} t} \ket{0}_a.
\label{eq:Supplement_sigmax1sigmaxn}
}
We remark that similar quantities have been studied in \cite{Supp_Igloi2000}. To evaluate the RHS of \eqref{eq:Supplement_sigmax1sigmaxn}, we use that $\mathcal{H}_{I}$ commutes with $e^{i \pi \sum_{k=1}^N c_k^\dagger c_k}$ for all $\xi$. Consequently, $\ket{0}_a$ is also an eigenstate of the latter operator \cite{Supp_Lieb1961}. It is now easy to see that
\ma{
\sigma_x^N  (t) \ket{0}_a &= [c_N^\dagger (t)- c_N (t)] e^{i \pi \sum_{k=1}^N c_k^\dagger c_k} \ket{0}_a\\
&= [c_N^\dagger (t)- c_N (t)] \ket{0}_a,
}
where $\mathcal{O}(t)= e^{i \mathcal{H}_{I,b} t} \mathcal{O} e^{-i \mathcal{H}_{I,b}
t}  $ for an operator $\mathcal{O}$. The same strategy as for $\langle \sigma_z^j
\rangle$ leads to 
\ma{
\langle &\sigma_x^1 \sigma_x^N \rangle (t) = \sum_k  \phi_{k,1}^b \psi_{k,N}^b + 2 \sum_{k,k^\prime} \lbrace \phi_{k,1}^b \psi_{k^\prime,N}^b  \times \notag \\
&[
X_{k,k\prime} \cos t(\Lambda_k^b + \Lambda_{k^\prime}^b) - 
Y_{k,k\prime} \cos t(\Lambda_k^b - \Lambda_{k^\prime}^b)
] \big\rbrace ,
}
with $X_{k,k\prime}$ and $Y_{k,k\prime}$ defined above. This result is plotted in
Fig.\ \ref{fig:S9} for a quench within the paramagnetic phase. The observable
$\langle \sigma_x^1 \sigma_x^N \rangle $ is an order parameter of the Ising chain in
equilibrium and does not develop a nonzero mean value for quenches within the
paramagnetic phase. However, at $t \approx N/2v_0=T/2$, where $v_0 = 2J$ in the
paramagnetic phase, oscillations of $\langle \sigma_x^1 \sigma_x^N \rangle $ arise.
After an abrupt increase, their amplitude decreases again, and this pattern
quasiperiodically repeats with period $T=N/v_0$. The observed behavior of the
end-to-end correlator can, again, be understood in the QP picture (see
\cite{Supp_Igloi2000} for a related analysis). Among the pairs of momentum-inverted QP
trajectories with the same origin only those originating at $j=N/2$ have trajectories
hitting the system boundaries simultaneously. Since only contiguously generated QPs
carry quantum correlations, only the QPs generated at $j=N/2$ can build up
correlations between the surface spins which will manifest themselves in a nonzero
value of $\langle \sigma_x^1 \sigma_x^N \rangle $. These QPs arrive for the first
time at the surface spins at $t=N/2v_0=T/2$, are then reflected, and build up
correlations between the surface spins each time they have travelled through the
whole chain, that is, after multiples of $T=N/v_0$. This explains the two different
time scales $T/2$ and $T$ and the quasiperiodicity (for $t>T/2$) of the end-to-end
correlator $\langle \sigma_x^1 \sigma_x^N \rangle (t)$. Slower QPs generated at
$j=N/2$ will also arrive simultaneously but delayed at the surface spins. They are
responsible for the slow decay of the oscillations of the correlator for $t \gtrsim
T/2 $ and will, for large $t$, eventually smear out the quasiperiodic structure.

\end{document}